\documentclass[preprint]{aastex}
\usepackage{epsfig}
\usepackage{pgfplots}
\usepackage{graphicx}
\usepackage{lscape}
\usepackage{amsmath}
\usepackage{amsbsy} 
\usepackage{rotating}
\usepackage{booktabs}
\usepackage{array}
\usepackage{lscape}
\usepackage[varg]{txfonts}
\usepackage{url}
\usepackage[colorlinks=true, linkcolor=blue, urlcolor=red]{hyperref}
\usepackage{breakurl}
\usepackage{float}

\begin{document}
\sloppy
\title{GRB 140606B / iPTF14bfu: Detection of shock-breakout emission from a cosmological $\gamma$-ray burst?}

\author{Z. Cano}
\affil{Centre for Astrophysics and Cosmology, Science Institute, University of Iceland, Reykjavik, Iceland.}
\email{zewcano@gmail.com}

\author{A. de Ugarte Postigo} 
\affil{Instituto de Astrof\'isica de Andaluc\'ia (IAA-CSIC), Glorieta de la Astronom\'ia s/n, E-18008, Granada, Spain.}
\affil{Dark Cosmology Centre, Niels Bohr Institute, Juliane Maries Vej 30, Copenhagen \O, D-2100, Denmark.}

\author{D. Perley} 
\affil{Cahill Center for Astrophysics, California Institute of Technology, Pasadena, CA 91125, USA.}

\author{T. Kr\"uhler} 
\affil{Dark Cosmology Centre, Niels Bohr Institute, Juliane Maries Vej 30, Copenhagen \O, D-2100, Denmark.}
\affil{European Southern Observatory, Alonso de C\'ordova 3107, Vitacura, Casilla 19001, Santiago 19, Chile.}

\author{R. Margutti} 
\affil{Harvard-Smithsonian Center for Astrophysics, 60 Garden St., Cambridge, MA 02138, USA.}

\author{M. Friis} 
\affil{Centre for Astrophysics and Cosmology, Science Institute, University of Iceland, Reykjavik, Iceland.}

\author{D. Malesani}
\affil{Dark Cosmology Centre, Niels Bohr Institute, Juliane Maries Vej 30, Copenhagen \O, D-2100, Denmark.}

\author{P. Jakobsson} 
\affil{Centre for Astrophysics and Cosmology, Science Institute, University of Iceland, Reykjavik, Iceland.}

\author{J. P. U. Fynbo}
\affil{Dark Cosmology Centre, Niels Bohr Institute, Juliane Maries Vej 30, Copenhagen \O, D-2100, Denmark.}

\author{J.~Gorosabel} 
\affil{Instituto de Astrof\'isica de Andaluc\'ia (IAA-CSIC), Glorieta de la Astronom\'ia s/n, E-18008, Granada, Spain.}
\affil{Unidad Asociada Grupo Ciencia Planetarias UPV/EHU-IAA/CSIC, Departamento de F\'isica Aplicada I, E.T.S. Ingenier\'ia, Universidad del Pa\'is-Vasco \\
UPV/EHU, Alameda de Urquijo s/n, E-48013 Bilbao, Spain.}
\affil{Ikerbasque, Basque Foundation for Science, Alameda de Urquijo 36-5, E-48008 Bilbao, Spain.}

\author{J. Hjorth}
\affil{Dark Cosmology Centre, Niels Bohr Institute, Juliane Maries Vej 30, Copenhagen \O, D-2100, Denmark.}

\author{R. S\'anchez-Ram\'irez}  
\affil{Instituto de Astrof\'isica de Andaluc\'ia (IAA-CSIC), Glorieta de la Astronom\'ia s/n, E-18008, Granada, Spain.}
\affil{Unidad Asociada Grupo Ciencia Planetarias UPV/EHU-IAA/CSIC, Departamento de F\'isica Aplicada I, E.T.S. Ingenier\'ia, Universidad del Pa\'is-Vasco \\
UPV/EHU, Alameda de Urquijo s/n, E-48013 Bilbao, Spain.}
\affil{Ikerbasque, Basque Foundation for Science, Alameda de Urquijo 36-5, E-48008 Bilbao, Spain.}

\author{S. Schulze} 
\affil{Instituto de Astrof\'isica, Facultad de F\'isica, Pontificia Universidad Cat\'olica de Chile, Vicu\~{n}a Mackenna 4860, 7820436 Macul, Santiago, Chile.}
\affil{Millennium Institute of Astrophysics, Vicu\~{n}a Mackenna 4860, 7820436 Macul, Santiago, Chile.}

\author{N. R. Tanvir} 
\affil{Department of Physics and Astronomy, University of Leicester, Leicester LE1 7RH, UK.}

\author{C. C. Th\"one}
\affil{Instituto de Astrof\'isica de Andaluc\'ia (IAA-CSIC), Glorieta de la Astronom\'ia s/n, E-18008, Granada, Spain.}

\author{D. Xu}
\affil{National Astronomical Observatories, Chinese Academy of Sciences, Beijing 100012, China.}
\affil{Key Laboratory of Space Astronomy and Technology, National Astronomical Observatories, Chinese Academy of Sciences, Beijing 100012, China.}
\affil{Dark Cosmology Centre, Niels Bohr Institute, Juliane Maries Vej 30, Copenhagen \O, D-2100, Denmark.}

\label{firstpage}

\begin{abstract}

We present optical and near-infrared photometry of GRB~140606B ($z=0.384$), and optical photometry and spectroscopy of its associated supernova (SN). The results of our modelling indicate that the bolometric properties of the SN ($M_{\rm Ni} = 0.4\pm0.2$~M$_{\odot}$, $M_{\rm ej} = 5\pm2$~M$_{\odot}$, and $E_{\rm K} = 2\pm1 \times 10^{52}$ erg) are fully consistent with the statistical averages determined for other GRB-SNe. However, in terms of its $\gamma$-ray emission, GRB~140606B is an outlier of the Amati relation, and occupies the same region as low-luminosity ($ll$) and short GRBs. The $\gamma$-ray emission in $ll$GRBs is thought to arise in some or all events from a shock-breakout (SBO), rather than from a jet. The measured peak photon energy ($E_{\rm p}\approx800$ keV) is close to that expected for $\gamma$-rays created by a SBO ($\gtrsim1$ MeV). Moreover, based on its position in the $M_{V,\rm p}$--$L_{\rm iso,\gamma}$~plane and the $E_{\rm K}$--$\Gamma\beta$~plane, GRB~140606B has properties similar to both SBO-GRBs and jetted-GRBs. Additionally, we searched for correlations between the isotropic $\gamma$-ray emission and the bolometric properties of a sample of GRB-SNe, finding that no statistically significant correlation is present. The average kinetic energy of the sample is $\bar{E}_{\rm K} = 2.1\times10^{52}$ erg. All of the GRB-SNe in our sample, with the exception of SN 2006aj, are within this range, which has implications for the total energy budget available to power both the relativistic and non-relativistic components in a GRB-SN event. 
\end{abstract}

\section{Introduction}

Supernova-like transients have now been observed to occur at the same spatial locations of both long- and short-duration $\gamma$-ray bursts (L/SGRBs).  The most common occurrences are of bright and energetic broad-lined Ic (IcBL) supernovae (SNe) that accompany LGRBs.  The first association was between GRB~980425 and SN~1998bw (Galama et al. 1998; Patat et al. 2001), and later examples include GRB~030329 and SN~2003dh (Hjorth et al. 2003; Stanek et al. 2003; Matheson et al. 2003), GRB~031203 and SN~2003lw (Malesani et al. 2004), GRB~060218 and SN~2006aj (Pian et al. 2006; Mazzali et al. 2006), GRB~100316D and SN~2010bh (Starling et al. 2011; Cano et al. 2011a, Olivares et al. 2012; Bufano et al. 2012), GRB~120422A and SN~2012bz (Melandri et al. 2012; Schulze et al. 2014), and GRB~130427A and SN~2013cq (Xu et al. 2013; Levan et al. 2014).  This increasing list of events has thoroughly strengthened the GRB-SN connection (Woosley \& Bloom 2006; Hjorth \& Bloom 2012; Cano 2013 -- C13 hereafter), and put their massive-star origins beyond any reasonable doubt.  On the flip-side, recently an $r$-process SN, also referred to as a ``mini-nova'' (Li \& Paczy\'nski 1998) or kilonova, was likely observed to accompany SGRB 130603B (Tanvir et al. 2013; Berger et al. 2013).  

Despite this common thread linking these two types of GRBs, their respective origins are distinctly different. LGRBs arise from the core-collapse of a massive star whose outer layers of hydrogen and helium have been stripped away prior to explosion, and whose stellar cores possess a large amount of angular momentum at the time of collapse -- angular momentum that is vital for eventually producing the observed $\gamma$-ray emission.  Conversely, an SGRB likely occurs during the merger of a binary compact object system, either a neutron star binary or a neutron star-black hole binary system.  

In both events an accretion disk is thought to form, which leads to the production of a relativistic bi-polar jet.  In the standard fireball model, shells of material within the jet interact producing the initial burst of $\gamma$-rays, called the prompt emission, via internal shocks.  As the jet propagates away from the explosion site, it eventually collides with the surrounding medium producing external shocks that power an afterglow (AG) that is visible across almost the entire EM spectrum, from X-rays to radio, and which lasts for severals weeks to months.  In this leptonic model, the prompt and AG radiation is synchrotron or synchrotron-self-Compton in origin (e.g. Rees \& M\'esz\'aros 1994).  However, the internal-shock model suffers from an inability to explain many features of prompt emission and AG light-curves (LCs).  Instead, alternative models have been proposed, such as the photospheric and hadronic emission models (e.g. Toma et al. 2011).  Photospheric models assume that thermal energy stored in the jet is radiated as prompt emission at the Thomson photosphere (Paczynski 1986; Thompson 1994; M\'esz\'aros \& Rees 2000).  Here the thermal energy in the jet can be produced by the dissipation of energy contained in the magnetic-field or from the particles themselves.  In hadronic models, synchrotron and inverse-Compton emission is produced by accelerated protons and secondary particles induced by the photo-pion cascade process (e.g. Vietri 1997; Asano et al. 2009).


The physical processes that power a GRB-SN arise via thermal heating from radioactive material trapped in the ejecta.  During the explosion radioactive nickel and cobalt are synthesized either by the neutrino wind emitted by the accretion disk (MacFayden \& Woosley 1999; MacFayden et al. 2001), or by a cocoon of material that surrounds the jet as it pierces through the progenitor star (e.g. Nagataki et al. 2006; Lazzati et al. 2012).  In LGRBs, typically 2--8 M$_{\odot}$ of material is ejected, of which 0.1--0.5 M$_{\odot}$ is in the form of radioactive $^{56}$Ni (e.g. C13).


\begin{figure*}
 \centering
 \includegraphics[bb=0 0 260 198, scale=1.6]{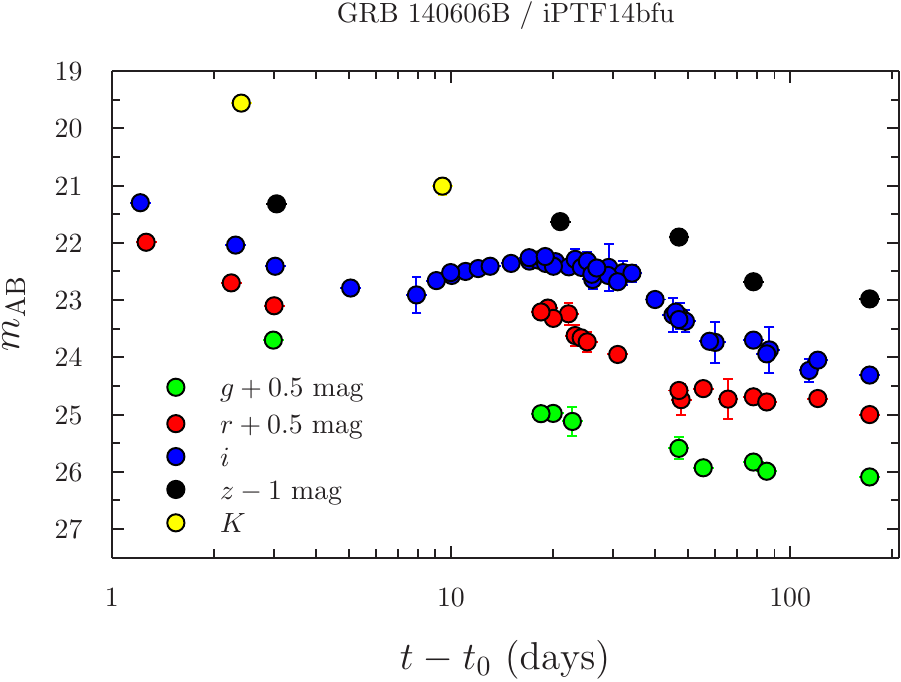}
 \caption{Observer-frame optical and NIR LCs of GRB~140606B~/~iPTF14bfu.  The apparent magnitudes in each filter are not corrected for foreground or rest-frame extinction, they are not host-subtracted, and times are observer-frame.  Early on, the LCs are powered by radiation from the forward-shock, which decay with a PL-like behaviour.  After 5--10 days the LC then starts to be dominated by flux emitted by the accompanying SN, which then too fades away.  By +171 days the only light detected is attributed as coming from the underlying host galaxy.}
 \label{fig:LC_mags}
\end{figure*}

In this paper we focus on a single LGRB event (GRB hereafter): GRB~140606B and its spectroscopically associated type Ic-BL SN.  GRB~140606B was detected at 03:11:51.86 UT on 06-June-2014 by the Fermi Gamma-Ray Burst Monitor (GBM), where a single, sharp pulse with a noisy tail was detected (Burns 2014).  Its duration\footnote{As presented in the on-line catalog at \url{http://heasarc.gsfc.nasa.gov/db-perl/W3Browse/w3query.pl}} was $T_{90} = 22.78\pm2.06$.  The GRB was also detected by Konus-Wind (Golenetskii et al. 2014) with a single pulse lasting $\sim 8$ s.  The GRB was not detected by the Burst Alert Telescope (BAT) aboard \emph{Swift} as it was outside of the spacecraft's field-of-view.  Contemporaneous observations were obtained with the Palomar 48 inch Oschin telescope (Singer et al. 2014), with several possible afterglow (AG) candidates to the GRB being identified.  Ultimately it was demonstrated that the X-ray AG of GRB~140606B (Mangano, Evans \& Goad 2014; Mangano \& Burrows) occurred within 1.9'' of the AG candidate iPTF14bfu.  The redshift of iPTF14bfu was measured to be $z=0.384$ by Perley et al. (2014), who also observed a nearby companion galaxy at the same redshift (see also Section \ref{sec:host_spectra}).  

The fluence detected by Fermi-GBM was $7.59\pm0.04 \times 10^{-6}$ erg cm$^{-2}$ in the 10--1000 keV (observer-frame) energy range.  The peak energy-cutoff of a band function fit to the $\gamma$-ray spectrum was $E_{\rm p} = 555\pm165$ keV, however, according to the on-line catalog, the Comptonized model provided the best fit to the $\gamma$-ray spectrum, where a cutoff energy of $E_{\rm p} = 579\pm135$ keV was determined.  The two energy-cutoffs are quite similar, and overlap in their respective error-bars.  We calculate\footnote{Using the spectral parameters presented in the on-line Fermi catalog and XSPEC.} a K-corrected, rest-frame isotropic energy release of in the 1--10,000 keV energy range of $E_{\rm iso} = (3.47\pm0.02) \times 10^{51}$ erg, and a rest-frame energy peak of $E_{\rm p} = 801\pm182$ keV. Further discussion of the prompt energetics are presented in Singer et al. (2015), and here in Section \ref{sec:SBO_vs_jet}.  The probability that GRB~140606B arose from a collapsar (Bromberg et al. 2013) based on the measurement of $T_{90}$ by GBM is $99\pm1\%$.  The basic observational properties of GRB~140606B / iPTF14bfu and its associated SN can be found in Table \ref{table:GRB_vitals}.

Throughout this paper we use a $\Lambda$CDM cosmology constrained by Planck (Planck Collaboration et al. 2013) of $H_{0} = 67.3$ km s$^{-1}$ Mpc$^{-1}$, $\Omega_{\rm M} = 0.315$, $\Omega_{\Lambda} = 0.685$.  Foreground extinction was calculated using the dust extinction maps of Schlegel et al. (1998) and Schlafly \& Finkbeiner (2011), where the value $E(B-V)_{\rm fore}=0.1022$ mag was used.  Unless stated otherwise, errors are statistical only.  Observer-frame times are used unless specified otherwise in the text.  The respective forward-shock afterglow decay and energy spectral indices $\alpha$ and $\beta$ are defined by $f_{\nu} \propto (t - t_{0})^{-\alpha}\nu^{-\beta}$, where $t_{0}$ is the time at which the GRB triggered the GBM instrument on-board the \emph{Fermi} satellite.

\begin{table*}
\centering
\setlength{\tabcolsep}{4.0pt}
\setlength{\extrarowheight}{3pt}
\caption{GRB 140606B / iPTF14bfu: vital statistics}
\label{table:GRB_vitals}
\begin{tabular}{rl}
\hline			
GRB 140606B / iPTF14bfu	&	Ref.	\\
\hline			
RA(J2000) = 21$^{h}$ 52$^{m}$ 29.97$^{s}$	&	Singer et al. (2014)	\\
Dec(J2000) = +32$^{d}$ 00$'$ 50.6$''$	&	Singer et al. (2014)	\\
$z=0.384$	&	Perley et al. (2014), this work	\\
$d_{\rm L}^{*} =  2144.4$ Mpc	&	this work	\\
$\mu^{*} = 41.66$ mag	&	this work	\\
$E(B-V)_{\rm fore} = 0.1022$ mag	&	Schlafly \& Finkbeiner (2011)	\\
$E(B-V)_{\rm rest} = 0.16\pm0.14$ mag	&	here	\\
$T_{90}$ (Fermi-GBM) = $22.78\pm2.06$ s	&	On-line Fermi catalog	\\
$E_{\rm \gamma, iso, rest} = (3.47\pm0.02) \times 10^{51}$ erg	&	On-line Fermi catalog; this work	\\
$E_{\rm \gamma, p,rest} = 801\pm182$ keV	&	On-line Fermi catalog; this work	\\
$v_{\rm ph, peak} = 19,820\pm1280$ km s$^{-1}$	&	this work, based on Fe \textsc{ii} $\lambda$5169	\\
$M_{\rm Ni} = 0.42\pm0.17$ M$_{\odot}$	&	this work	\\
$M_{\rm ej} = 4.8\pm1.9$ M$_{\odot}$	&	this work	\\
$E_{\rm K} = (1.9\pm1.1) \times 10^{52}$ erg	&	this work	\\
\hline					
\end{tabular}
\begin{flushleft}
$^{*}$ Calculated using $H_{0} = 67.3$ km s$^{-1}$ Mpc$^{-1}$, $\Omega_{M} = 0.315$, $\Omega_{\Lambda} = 0.685$. \\ 
\end{flushleft}
\end{table*}

\section{Data Acquisition \& Reduction}
\label{section:data}

\subsection{Photometry}
\label{sec:photometry}

We obtained observations with several ground-based telescopes: the 2 m Liverpool Telescope (LT), the 2.5 m Nordic Optical Telescope (NOT), and the 10.4 m Gran Telescopio Canarias (GTC) telescope, all in La Palma, Spain; the 10.0 m Keck I telescope on Mauna Kea, Hawaii; and the Palomar 60 inch (P60) telescope in San Diego County, California.  Nine epochs of $griz$ observations were obtained with the NOT; 13 epochs of observations were obtained with the LT, 12 being in $i$-band and the solitary other in $r$.  Ten epochs of $griz$ photometry were obtained with the GTC, including late-time images (+171 days) of just the host galaxy, which were used as templates for the image subtraction technique (see below) as well correcting for the host contribution in all earlier observations obtained on all telescopes (see Section \ref{sec:LC_decomposition}).  Twenty-six epochs of $gri$ photometry were obtained with the P60 (Cenko et al. 2006), some of which also appears in Singer et al. (2015).  Three epochs of photometry were obtained with Keck-LRIS (Oke et al. 1995) in filters $gRR_{s}i$, and two epochs of $K$-band images with Keck-MOSFIRE (McLean et al. 2012).  Late-time P60 images in $g$ (26, 28, 29, 30-June and 01-July) and $r$ (22, 24 \& 25-July into one image; and 05, 08, 10, 12 \& 16-August into another) were co-added to increase the S/N of the optical transient (OT).

Image reduction of photometric data obtained on all telescopes except those obtained with Keck and P60 was performed using standard techniques in IRAF\footnote{IRAF is distributed by the National Optical Astronomy Observatory, which is operated by the Association of Universities for Research in Astronomy, Inc., under cooperative agreement with the National Science Foundation.}.  Photometric reduction of the Keck and P60 data were performed using a combination of \textsc{python} and \textsc{idl} routines.  Observations of Landolt/Stetson standard field PG2213-006 (Landolt 1992; Stetson 2000) were obtained with the NOT on the night of 26--27$^{th}$-June-2014, along with images of the GRB, with both fields obtained in filters $griz$.  The $BVRI$ magnitudes of PG2213-006 were transformed into $griz$ using transformation equations from Lupton (2005)\footnote{http://www.sdss.org/dr4/algorithms/sdssUBVRITransform.html}, where the photometric uncertainties and rms scatter in the transformation equations were combined with the instrumental magnitude errors in quadrature.  Zeropoints between the instrumental and catalog magnitudes were determined for the standard field, and these were used to calibrate a set of almost 100 secondary standard stars in the GRB FOV.  All images on all telescopes were then calibrated to these secondary standards via a zeropoint, except for the $R$ and $R_{s}$ images obtained with Keck-LRIS, which were calibrated with a zeropoint and a colour term using the $g$-band images to determine $g-R$ and $g-R_{s}$.

We used our deep GTC images to obtain image-subtracted magnitudes of the optical transient (OT) associated with GRB~140606B, using the final epoch in each filter as a template.    Image subtraction was performed using an adaptation of the original ISIS program (Alard \& Lupton 1998; Alard 2000) that was developed for Hubble Space Telescope SN surveys by Strolger et al. (2004).  A key advantage of this code is the option for the user to specify a set of stamps for the program to use when it calculates the point-spread function in each image.  The image-subtraction technique was then optimized by varying the kernel mesh size and measuring the standard deviation ($\sigma$) of the background counts in a nearby region in the image (where images with lower $\sigma$ values indicate that they are a better subtracted image).  As a self-consistency check, we compared the OT magnitudes against those found by performing photometry on the un-subtracted images, converting the magnitudes into fluxes, and then mathematically subtracting the host flux.  Good agreement was obtained with both methods, showing that the image-subtraction technique was well optimized.

The $griz$ magnitudes of the host galaxy were measured, and these magnitudes were converted into monochromatic fluxes using the flux zeropoints from Fukugita et al. (1995), and then subtracted from the earlier observations (also converted into monochromatic fluxes) obtained with the other instruments.  The apparent magnitudes (not corrected for foreground or host extinction) of the GRB+SN+host are presented in Table \ref{table:photometry_obs_log}, and GTC photometry of the GRB's host galaxy and nearby companion (see Section \ref{sec:host_spectra}) are presented in Table \ref{table:photometry_HOST_obs_log}.  All magnitudes are in the AB system, except the $K$-band observations, which are Vega.

\subsection{Spectroscopy}

We obtained seven epochs of spectroscopy of GRB~140606B and its accompanying SN, see Table \ref{table:spectra_obs_log}.  Four epochs were obtained with GTC-OSIRIS, all with the R500R grism that has a spectral resolution of $\delta\lambda/\lambda\sim600$ and coverage from 4800 to 10000 \AA.  All epochs consisted of $3\times1200$ s exposure times.  We also obtained three epochs of spectroscopy with Keck-LRIS with the 600/4000 (blue) grism and the 400/8500 (red) grating, which cover a total wavelength range of 310-1030 nm.  The GTC spectra were reduced using standard techniques with IRAF-based scripts, while the Keck spectra were reduced using \textsc{idl} routines.  The final epoch of GTC spectra obtained at +171 days of the host and nearby companion galaxy were each flux calibrated using their contemporaneous GTC $griz$ photometry.  An analysis of the GRB host galaxy and its companion is presented in Section \ref{sec:host_spectra}.

\begin{table}
\scriptsize
\centering
\setlength{\tabcolsep}{6.0pt}
\setlength{\extrarowheight}{3pt}
\caption{GRB 140606B / iPTF14bfu: Photometry observation log}
\label{table:photometry_obs_log}
\begin{tabular}{cccccccc}
\hline																														
$t-t_{0}$ (days)				&	Filter	&		mag$^{a}$				&	Telescope	&		$t-t_{0}$ (days)				&	Filter	&		mag$^{a}$				&	Telescope	\\
\hline																														
$	2.994	\pm	0.000	$	&	$g$	&	$	23.20	\pm	0.05	$	&	NOT	&	$	17.014	\pm	0.000	$	&	$i$	&	$	22.26	\pm	0.09	$	&	GTC	\\
$	18.435	\pm	0.000	$	&	$g$	&	$	24.49	\pm	0.03	$	&	Keck	&	$	17.054	\pm	0.000	$	&	$i$	&	$	22.32	\pm	0.06	$	&	LT	\\
$	20.024	\pm	0.000	$	&	$g$	&	$	24.48	\pm	0.09	$	&	NOT	&	$	18.026	\pm	0.000	$	&	$i$	&	$	22.30	\pm	0.06	$	&	LT	\\
$	22.812	\pm	2.467	$	&	$g$	&	$	24.62	\pm	0.25	$	&	P60	&	$	18.972	\pm	0.000	$	&	$i$	&	$	22.24	\pm	0.05	$	&	GTC	\\
$	47.007	\pm	0.000	$	&	$g$	&	$	25.02	\pm	0.19	$	&	NOT	&	$	19.011	\pm	0.000	$	&	$i$	&	$	22.36	\pm	0.06	$	&	LT	\\
$	55.487	\pm	0.000	$	&	$g$	&	$	25.43	\pm	0.09	$	&	Keck	&	$	20.058	\pm	0.000	$	&	$i$	&	$	22.41	\pm	0.04	$	&	NOT	\\
$	77.981	\pm	0.000	$	&	$g$	&	$	25.27	\pm	0.09	$	&	GTC	&	$	20.331	\pm	0.000	$	&	$i$	&	$	22.33	\pm	0.12	$	&	P60	\\
$	85.376	\pm	0.000	$	&	$g$	&	$	25.49	\pm	0.12	$	&	Keck	&	$	21.297	\pm	0.000	$	&	$i$	&	$	22.58	\pm	0.24	$	&	P60	\\
$	1.262	\pm	0.000	$	&	$r$	&	$	21.49	\pm	0.08	$	&	P60	&	$	22.259	\pm	0.000	$	&	$i$	&	$	22.42	\pm	0.12	$	&	P60	\\
$	2.250	\pm	0.000	$	&	$r$	&	$	22.20	\pm	0.10	$	&	P60	&	$	23.261	\pm	0.000	$	&	$i$	&	$	22.29	\pm	0.18	$	&	P60	\\
$	3.010	\pm	0.000	$	&	$r$	&	$	22.60	\pm	0.04	$	&	NOT	&	$	24.246	\pm	0.000	$	&	$i$	&	$	22.43	\pm	0.10	$	&	P60	\\
$	18.434	\pm	0.000	$	&	$r$	&	$	22.71	\pm	0.03	$	&	Keck	&	$	25.256	\pm	0.000	$	&	$i$	&	$	22.32	\pm	0.16	$	&	P60	\\
$	19.316	\pm	0.000	$	&	$r$	&	$	22.64	\pm	0.09	$	&	P60	&	$	26.032	\pm	0.000	$	&	$i$	&	$	22.55	\pm	0.08	$	&	NOT	\\
$	20.042	\pm	0.000	$	&	$r$	&	$	22.82	\pm	0.05	$	&	NOT	&	$	26.927	\pm	0.000	$	&	$i$	&	$	22.44	\pm	0.11	$	&	GTC	\\
$	22.220	\pm	0.000	$	&	$r$	&	$	22.74	\pm	0.19	$	&	P60	&	$	29.061	\pm	0.000	$	&	$i$	&	$	22.57	\pm	0.06	$	&	LT	\\
$	23.248	\pm	0.000	$	&	$r$	&	$	23.12	\pm	0.19	$	&	P60	&	$	31.024	\pm	0.000	$	&	$i$	&	$	22.68	\pm	0.05	$	&	NOT	\\
$	24.211	\pm	0.000	$	&	$r$	&	$	23.16	\pm	0.12	$	&	P60	&	$	31.184	\pm	0.000	$	&	$i$	&	$	22.65	\pm	0.16	$	&	P60	\\
$	25.227	\pm	0.000	$	&	$r$	&	$	23.23	\pm	0.17	$	&	P60	&	$	32.180	\pm	0.000	$	&	$i$	&	$	22.52	\pm	0.21	$	&	P60	\\
$	31.042	\pm	0.000	$	&	$r$	&	$	23.45	\pm	0.07	$	&	NOT	&	$	34.184	\pm	0.000	$	&	$i$	&	$	22.53	\pm	0.15	$	&	P60	\\
$	47.015	\pm	0.000	$	&	$r$	&	$	24.08	\pm	0.13	$	&	NOT	&	$	40.007	\pm	0.000	$	&	$i$	&	$	22.99	\pm	0.16	$	&	LT	\\
$	47.745	\pm	1.496	$	&	$r$	&	$	24.24	\pm	0.27	$	&	P60	&	$	45.151	\pm	0.000	$	&	$i$	&	$	23.26	\pm	0.29	$	&	P60	\\
$	55.487	\pm	0.000	$	&	$r$	&	$	24.05	\pm	0.08	$	&	Keck	&	$	46.015	\pm	0.000	$	&	$i$	&	$	23.10	\pm	0.10	$	&	LT	\\
$	65.657	\pm	5.428	$	&	$r$	&	$	24.23	\pm	0.35	$	&	P60	&	$	47.025	\pm	0.000	$	&	$i$	&	$	23.34	\pm	0.08	$	&	NOT	\\
$	77.990	\pm	0.000	$	&	$r$	&	$	24.05	\pm	0.07	$	&	GTC	&	$	47.309	\pm	0.000	$	&	$i$	&	$	23.28	\pm	0.23	$	&	P60	\\
$	85.386	\pm	0.000	$	&	$r$	&	$	24.28	\pm	0.08	$	&	Keck	&	$	49.175	\pm	0.000	$	&	$i$	&	$	23.37	\pm	0.19	$	&	P60	\\
$	120.764	\pm	0.000	$	&	$r$	&	$	24.22	\pm	0.11	$	&	GTC	&	$	57.901	\pm	0.000	$	&	$i$	&	$	23.72	\pm	0.10	$	&	NOT	\\
$	1.213	\pm	0.000	$	&	$i$	&	$	21.30	\pm	0.11	$	&	P60	&	$	60.154	\pm	0.000	$	&	$i$	&	$	23.74	\pm	0.36	$	&	P60	\\
$	2.317	\pm	0.000	$	&	$i$	&	$	22.04	\pm	0.12	$	&	P60	&	$	78.000	\pm	0.000	$	&	$i$	&	$	23.70	\pm	0.07	$	&	GTC	\\
$	3.031	\pm	0.000	$	&	$i$	&	$	22.41	\pm	0.06	$	&	NOT	&	$	85.373	\pm	0.000	$	&	$i$	&	$	23.94	\pm	0.09	$	&	Keck	\\
$	5.059	\pm	0.000	$	&	$i$	&	$	22.79	\pm	0.10	$	&	LT	&	$	86.948	\pm	0.000	$	&	$i$	&	$	23.87	\pm	0.40	$	&	GTC	\\
$	8.045	\pm	0.000	$	&	$i$	&	$	22.87	\pm	0.53	$	&	GTC	&	$	113.844	\pm	0.000	$	&	$i$	&	$	24.23	\pm	0.20	$	&	NOT	\\
$	9.058	\pm	0.000	$	&	$i$	&	$	22.66	\pm	0.11	$	&	GTC	&	$	120.746	\pm	0.000	$	&	$i$	&	$	24.05	\pm	0.08	$	&	GTC	\\
$	9.984	\pm	0.000	$	&	$i$	&	$	22.52	\pm	0.12	$	&	GTC	&	$	3.061	\pm	0.000	$	&	$z$	&	$	22.32	\pm	0.10	$	&	NOT	\\
$	10.053	\pm	0.000	$	&	$i$	&	$	22.57	\pm	0.11	$	&	LT	&	$	21.004	\pm	0.000	$	&	$z$	&	$	22.63	\pm	0.10	$	&	NOT	\\
$	11.045	\pm	0.000	$	&	$i$	&	$	22.50	\pm	0.08	$	&	LT	&	$	47.049	\pm	0.000	$	&	$z$	&	$	22.90	\pm	0.12	$	&	NOT	\\
$	12.032	\pm	0.000	$	&	$i$	&	$	22.45	\pm	0.08	$	&	LT	&	$	78.013	\pm	0.000	$	&	$z$	&	$	23.68	\pm	0.10	$	&	GTC	\\
$	13.053	\pm	0.000	$	&	$i$	&	$	22.41	\pm	0.07	$	&	LT	&	$	2.408	\pm	0.000	$	&	$K$	&	$	19.56	\pm	0.12	$	&	Keck	\\
$	15.048	\pm	0.000	$	&	$i$	&	$	22.36	\pm	0.06	$	&	LT	&	$	9.440	\pm	0.000	$	&	$K$	&	$	21.01	\pm	0.15	$	&	Keck	\\
\hline
\end{tabular}
\begin{flushleft}
$^{a}$ Apparent magnitudes, which are not corrected for foreground or rest-frame extinction. \\
NB: All magnitudes are in the AB system, except the $K$-band observations, which are Vega.\\
NB: Host photometry can be found in Table \ref{table:photometry_HOST_obs_log}.\\
NB: Error-bars given for specific $t-t_{0}$ values arise from images that were co-added from several iPTF epochs in order to increase the S/N of the data.\\
\end{flushleft}
\end{table}

\begin{table*}
\scriptsize
\centering
\setlength{\tabcolsep}{6.0pt}
\setlength{\extrarowheight}{3pt}
\caption{GRB 140606B / iPTF14bfu: Spectroscopy observation log}
\label{table:spectra_obs_log}
\begin{tabular}{ccccccc}
\hline													
UT date	&	UT time$^{a}$	&	$t-t_{0}$ (d)	&	Phase$^{b}$	&	Range (\AA)	&	Equipment	&	Exposure Time	\\
\hline													
24-Jun-2014	&	13:00:17	&	18.4089	&	-1.0	&	$3100-10300$	&	Keck, LRIS,  600/4000 (blue)  \& 400/8500 (red) 	&	1200 s	\\
25-Jun-2014	&	02:44:40	&	18.9824	&	-0.4	&	$4800-10000$	&	GTC, Osiris, R500R 	&	3$\times$1200 s	\\
29-Jun-2014	&	13:13:28	&	23.4187	&	4.0	&	$3100-10300$	&	Keck, LRIS,  600/4000 (blue)  \& 400/8500 (red) 	&	3600 s	\\
03-Jul-2014	&	01:32:44	&	26.9325	&	7.5	&	$4800-10000$	&	GTC, Osiris, R500R 	&	3$\times$1200 s	\\
30-Jul-2014	&	14:17:36	&	55.4626	&	36.1	&	$3100-10300$	&	Keck, LRIS,  600/4000 (blue)  \& 400/8500 (red) 	&	900 s	\\
01-Sept-2014	&	02:00:28	&	86.9517	&	67.6	&	$4800-10000$	&	GTC, Osiris, R500R 	&	3$\times$1200 s	\\
25-Nov-2014	&	20:12:24	&	172.7100	&	153.3	&	$4800-10000$	&	GTC, Osiris, R500R 	&	3$\times$1200 s	\\
\hline	
\end{tabular}
\begin{flushleft}
$^{a}$ UT start time. \\
$^{b}$ Relative to peak, observer-frame $i$-band light. \\
\end{flushleft}
\end{table*}

\section{Light Curve Analysis}
\label{section:LC_analysis}

Figure \ref{fig:LC_mags} presents our optical and NIR photometry of GRB~140606B / iPTF14bfu.  An initial power-law (PL)-like decay is seen in filters $r$ and $i$, after which light from the accompanying SN becomes the dominant source of flux, before it too faded into obscurity.  The final epoch (+171 days, observer-frame) in each filter corresponds to light coming from just the host galaxy.  The magnitudes in Fig. \ref{fig:LC_mags} are not corrected for foreground or rest-frame extinction.

\subsection{Decomposing the optical light curves}
\label{sec:LC_decomposition}

\begin{figure*}
 \centering
 \includegraphics[bb=0 0 1379 836, scale=0.3]{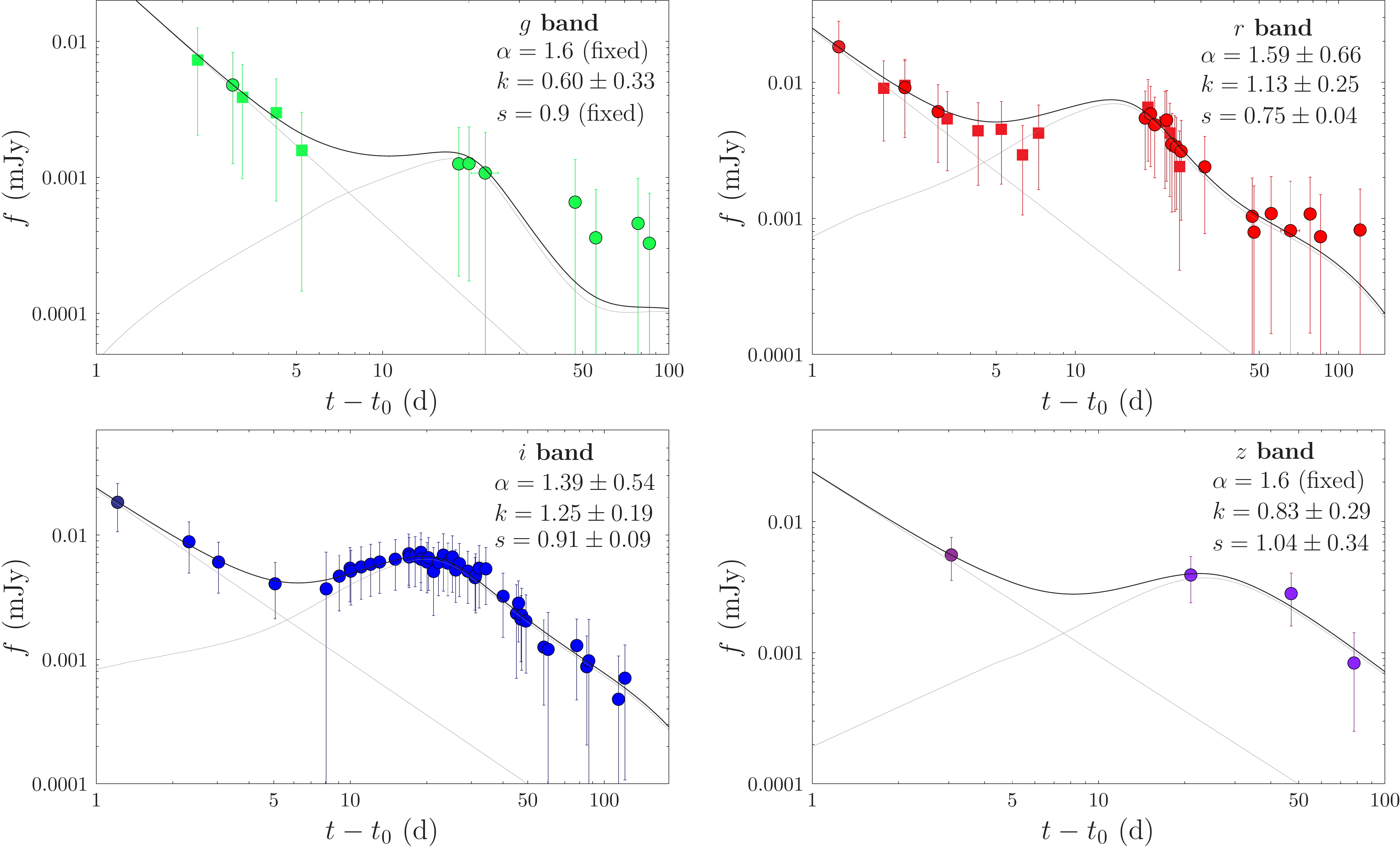}
 \caption{Decomposed flux LCs of GRB 140606B / iPTF14bfu.  The magnitudes presented in Fig. \ref{fig:LC_mags} have been corrected for foreground and rest-frame extinction, converted into monochromatic fluxes, and the host contribution in each filter has been mathematically subtracted away.  Optical observations in $g$ and $r$ from Singer et al. (2015) are shown as filled squares in each subplot, and have been included to assist with the LC modelling.  The time axis is in the observer frame.  Presented in each subplot are the best-fitting parameters of the fitted model, with consists of a single power-law and a luminosity ($k$) and time ($s$) stretched SN template.  The values of $k$ and $s$ are relative to SN~1998bw if it occurred at $z=0.384$ in each given observer-frame filter. }
 \label{fig:LC_decomp}
\end{figure*}

For every cosmological GRB-SN event, light arises from three sources (Zeh et al. 2004; Ferrero et al. 2006; Cano et al. 2011b; Hjorth 2013): the optical afterglow (AG), the accompanying SN, and a constant source of flux from the underlying host galaxy.  In this respect GRB 140606B / iPTF14bfu is no different.  The first step towards isolating the SN properties is to account for, then remove, the AG and host contributions.  

Before the host flux was subtracted away, all optical photometry, including that of the host galaxy, were corrected for foreground extinction\footnote{Note that this applies only to images that were not image-subtracted -- i.e. all images apart from the GTC ones.  For the GTC images, correction for foreground extinction was performed after the image-subtraction was performed.}.  Next, all magnitudes were converted into monochromatic fluxes using the flux zeropoints from Fukugita et al. (1995) and Cohen et al. (2003).  Then, in all relevant filters the host flux was mathematically subtracted away, leaving light from just the AG and the SN.  At this point the LCs were fitted with an analytical model that considers light from synchrotron radiation, which has a PL-like temporal behaviour, and a template SN, which was SN~1998bw.  Template SN LCs (C13) are created using the original observations of SN~1998bw, which were well sampled across optical filters $UBVRIJH$ (Galama et al. 1998; Patat et al. 2001; Clocchiati et al. 2011).  An optical/NIR spectral energy distribution (SED) is created at each epoch of contemporaneous optical photometry, which is fit with a cubic spline and then interpolated to the desired redshifted rest-frame wavelength.  The template LC is then flux normalized to the chosen redshift using the ratio of luminosity distances calculated by the program itself.  Thus the resultant template LC is a K-corrected representation of how SN~1998bw would appear if it occurred at the redshift and observer-frame filter of the GRB-SN being modelled.  The template LC is then fit with a linear spline, and a final function that was the sum of the AG and SN components is fitted to the observations to determine the AG decay rate ($\alpha$), and the luminosity ($k$) and stretch ($s$) of the considered SN relative to the template (see equation 5 in C13).

The best-fitting parameters in filters $griz$ are presented in Table \ref{table:GRB_SN_obs_props}.  Due to the lack of optical data in most filters apart from $i$, we included the published photometry from Singer et al. (2015), which included useful $g$- and $r$-band data (shown as filled squares in Fig. \ref{fig:LC_decomp}).  In all filters a single power-law (SPL) described the data in all filters with no need to invoke the presence of a jet-break, which can also be seen visually in Figs. \ref{fig:LC_mags} and \ref{fig:LC_decomp}.  The value of $\alpha$ is approximately the same in all filters, where in $r$ and $i$ (which were the only datasets where there were enough datapoints that we could allow all of the parameters to vary freely) we obtained values of $\alpha=1.59\pm0.66$ and $\alpha=1.39\pm0.54$, respectively.  The value of $\alpha$ in these filters agree within their respective error-bars, and imply a decay rate of $\alpha_{\rm opt} \approx 1.5$.  The value of $\alpha$ determined from the $K$-band data is smaller ($\alpha\approx1$), though we must consider the caveat that the host and SN contributions have not yet been removed, and should any light from either of these two sources contribute a non-negligible contribution of flux to the LC at these epochs, the effect will be a LC that decays faster when these contributions are removed.

Despite the additional data from Singer et al. (2015) in the $g$-band there was still a paucity of data, and we had to fix the decay constant of the AG component (in the range $1.4 \le \alpha \le 1.9$, as well as the stretch factor ($0.5 \le k \le 1.0$), whose ranges were motivated by the values determined in the $r$- and $i$-bands, thus fitting for only $k$.  Additionally in the $z$-band, we also fixed the value of the decay constant to $1.4 \le \alpha \le 1.9$ due to the lack of early data, which allowed us to fit for both $k$ and $s$.  Note that the $k$ value of the $g$-band data are very tentative, and likely do no properly describe the SN associated with GRB~140606B, especially at late times.  When creating the synthetic SN~1998bw LCs, the SEDs of SN~1998bw range from rest-frame $U$ to $H$, where the effective wavelength of the $U$-band filter is $\lambda=3652$ \AA $ $ (Fukugita et al. 1995).  At $z=0.384$, observer-frame $g$-band corresponds to $4858/(1+z)=3510$ \AA, which is slightly bluer than the input SEDs, and therefore the SED interpolation provides only a loose approximation for the observer-frame $g$-band LC.  The template LCs in the other observer-frame filters are more trustworthy however, and do not suffer from this limitation.

Using the values of $s$ and $k$ from Table \ref{table:GRB_SN_obs_props}, we estimated the peak magnitudes and time of peak light in observer-frame filters $r$ and $i$.  It is seen that the SN peaks at a later time in the redder $i$-band filter relative to the $r$-band.  Using a distance modulus of $\mu=41.66$ mag (Table \ref{table:GRB_vitals}), in case (1) we found peak, observer-frame, absolute magnitudes of $M_{r, \rm obs} = -19.61\pm0.27$ and $M_{i, \rm obs} = -19.78\pm0.18$.

\begin{table}
\centering
\setlength{\tabcolsep}{2.0pt}
\setlength{\extrarowheight}{3pt}
\caption{GRB 140606B / iPTF14bfu: AG \& SN observational properties}
\label{table:GRB_SN_obs_props}
\begin{tabular}{cccccc}
\hline																					
Filter	&	$\alpha$	&	$k$	&	$s$	&	$m_{\rm p}$ (mag)	&	$t_{\rm p}$ (days)	\\
\hline																					
$g$	&	$(1.4-1.9)^{\dagger}$	&	$0.60\pm0.33$	&	$(0.5-1.0)^{\dagger}$	&	-	&	-	\\
$r$	&	$1.59\pm0.66$	&	$1.13\pm0.25$	&	$0.75\pm0.04$	&	$22.05\pm0.27$	&	$16.32\pm1.63$	\\
$i$	&	$1.39\pm0.54$	&	$1.25\pm0.19$	&	$0.91\pm0.09$	&	$21.88\pm0.18$	&	$20.17\pm2.61$	\\
$z$	&	$(1.4-1.9)^{\dagger}$	&	$0.83\pm0.29$	&	$1.04\pm0.34$	&	-	&	-	\\
$K$	&	$0.98\pm0.15$	&	-	&	-	&	-	&	-	\\
$V_{\rm rest}$	&	$1.56\pm0.35$	&	$1.04\pm0.24$	&	$0.81\pm0.13$	&	$22.08\pm0.28$	&	$17.67\pm3.11$	\\
\hline																						
\end{tabular}
\begin{flushleft}
$^{\dagger}$ Values are fixed during the fit. \\
NB: $K$-band data are not host or SN subtracted. \\
\end{flushleft}
\end{table}

\subsection{Color curves}
\label{sec:color_curves}

\begin{figure}
 \centering
 \includegraphics[bb=0 0 266 172, scale=1.5]{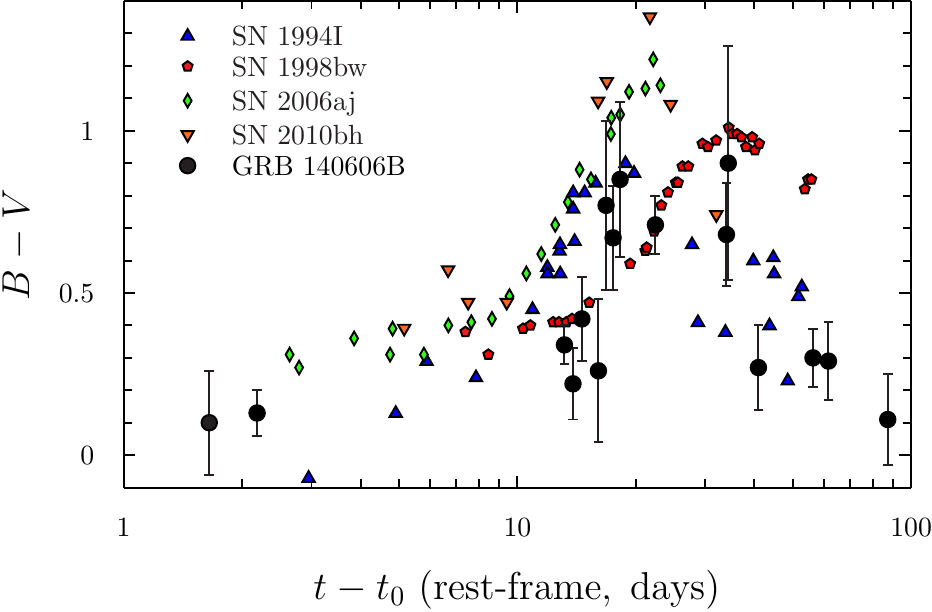}
 \caption{GRB 140606B / iPTF14bfu: Observer-frame $r-i$ (AG+SN) $vs$ rest-frame $B-V$ for a sample of GRB-SNe and SN Ic 1994I, which was not associated with a GRB.  At $z=0.384$, observer-frame $r$ and $i$ roughly correspond to rest frame $B$ and $V$, respectively.  It is seen that the colour curves of GRB~140606B evolve at a similar rate to SN~2006aj and SN~2010bh, but more rapidly than SN~1998bw, and slower than SN~1994I.  The peak colours for GRB~140606B are smaller than all of the other GRB-SNe, and reach similar peak values as seen for SN~1994I.  All times are given as rest-frame, and all events have been corrected for foreground and rest-frame extinction, apart from GRB~140606B which has only been corrected for the former.  See the main text for references to the different photometric datasets.}
 \label{fig:ri_colour}
\end{figure}

Observer-frame filters $r$ and $i$ roughly correspond to rest-frame filters $B$ and $V$, respectively.  Using the effective wavelengths for each of the aforementioned filters from Fukugita et al. (1995), observer-frame $r$ and $i$ correspond to rest-frame $6290/(1+z) = 4549$ \AA $ $ and $7706/(1+z) = 5568$ \AA, respectively.  The effective wavelengths of $B$ and $V$ are 4448 \AA $ $ and 5505 \AA, respectively.  The difference between $r_{\rm obs}$ and $B_{\rm rest}$ is 101 \AA, while the difference between $i_{\rm obs}$ and $V_{\rm rest}$ is 63 \AA.

Making the assumption that $(r-i)_{\rm obs} \approx (B-V)_{\rm rest}$, we plotted in Fig. \ref{fig:ri_colour} the former (host-subtracted) values in rest-frame times, where the errors have been added in quadrature, against $B-V$ for a set of SNe Ibc: SN~1994 (Richmond et al. 1996), SN~1998bw (Galama et al. 1998), SN~2006aj (Pian et al. 2006; Sollerman et al. 2006) and SN~2010bh (Cano et al. 2011b).  All but the first are GRB-SNe, while SN~1994I is a SN Ic not associated with a GRB.  Each of the comparison SNe have been corrected for observer-frame and rest-frame extinction using the values determined by the authors, and as tabulated in C13.  The colour $(r-i)_{\rm obs}$ for GRB~140606B seen in Fig. \ref{fig:ri_colour} was corrected for foreground extinction only as the increased error-bars of $r$ and $i$ from the loosely constrained rest-frame extinction make such a comparison quite uninformative.

It is seen that the temporal evolution of $B-V$ for GRB~140606B occurs on timescales similar to those seen for SNe 2006aj and 2010bh.  The evolution of GRB~140606B is more rapid that for SN~1998bw, but takes longer to reach peak values than seen for SN~1994I.  It is also seen that the peak colour for GRB~140606B does not get to as large values seen for the other GRB-SNe, and instead reaches peak values seen for SN~1994I.

\subsection{$k$--$s$ relation}
\label{sec:ks}

\begin{figure}
 \centering
 \includegraphics[bb=0 0 576 432, scale=0.8]{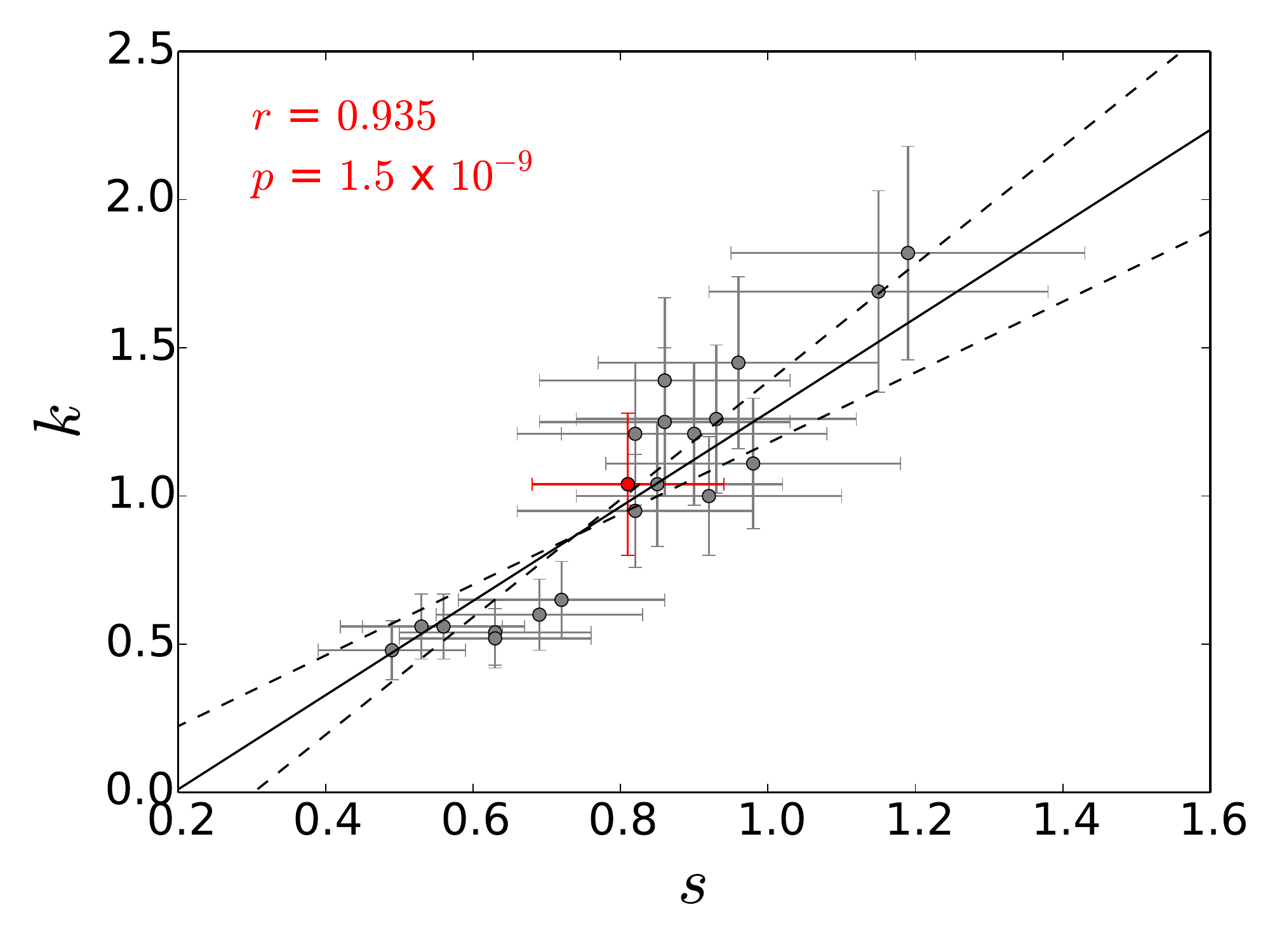}
 \caption{GRB 140606B / iPTF14bfu: $k$--$s$ relation.  Rest-frame $V$-band is shown in red, which has been corrected for foreground \& rest-frame extinction, as well as being host-subtracted.  Also plotted is the sample of GRB-SNe from C14 (grey).  A bootstrap analysis was conducted where the $k$ and $s$ values in $V$-band were added to the sample in C14, and a linear equation was fit to the combined dataset to determine the best-fitting values of the $y$-intercept ($b$), the slope ($m$), while the Pearson's correlation coefficient ($r$) and the two-point probability of a chance correlation ($p$) were also calculated.  We found $m=1.59\pm0.19$, $b=-0.30\pm0.14$, $r=0.935$ and $p=1.5\times10^{-9}$.}
 \label{fig:ks}
\end{figure}

It was shown in Cano (2014; C14 hereafter) that GRB-SN have a luminosity--stretch relation analogous to the luminosity--decline relation observed for SNe Ia in the early 1990s (Phillips 1993).  Such a relation can be explained physically primarily by the diffusion timescale, where for SNe of similar average opacities, in cases of SNe with more ejecta, trapped radiation will take longer to escape into space than for SNe that possess relatively less ejecta.  Moreover, as pointed out by Woosley et al. (2007), increased nickel content in the ejecta leads to larger temperatures and ionization levels, which further increase the opacity, and hence increase the diffusion timescale.  Furthermore, as discussed by Kasen \& Woosley (2007), dimmer SNe~Ia are relatively cooler than brighter SNe~Ia, which allows Fe \textsc{iii} to recombine more rapidly than in brighter/hotter events, thus providing additional line-blanketing/flux-suppressing effects that result in LCs that decay more rapidly.  Such line-blanketing effects from iron-group elements will undoubtedly affect the visual opacity of GRB-SNe, which also suffer opacity due to electron scattering.

In C14, optical fluxes at exact rest-frame filters were decomposed from the observer-frame observations of a sample of nine GRB-SNe using an analysis identical to that described in Section \ref{sec:LC_decomposition}.  Once the sample was created, a bootstrap analysis was performed to determine the $y$-intercept ($c$) and slope ($m$) of a linear equation fit to the dataset, as well as calculating the Pearson's correlation coefficient and the two-point probability of a chance correlation ($p$). It was found that a statistically significant correlation was seen in the $k$ and $s$ values for the nine GRB-SNe in rest-frame filters $UBVRI$.  As $k$ and $s$ are proxies for the luminosity and width of each GRB-SN in each filter (also used for SNe Ia, e.g. Perlmutter et al. 1997; Goldhaber et al. 2001), the correlation between them implies that like SNe Ia, GRB-SNe also have an analogous luminosity--decline relationship.  A similar conclusion was reached by Li \& Hjorth (2014) for a set of GRB-SNe, with many events overlapping between the two studies.

With each newly discovered and observed GRB-SN, there is a desire to understand if it too obeys the $k$--$s$ relation.  At $z=0.384$, rest-frame $V$-band occurs at observer-frame $\lambda_{V} = 5505(1+z) = 7618.9$ \AA, which falls between observer-frame $r$ and $i$, (e.g. see Section \ref{sec:color_curves}).  We then followed the procedure described in C14 to decompose and model the observations to determine $k$ and $s$ (which assumes a fix cosmology to create the template LCs).  When performing the modelling, we found that the temporal decay index was similar to that of observer-frame $i$ (unsurprisingly as rest-frame $V$ is redshifted to almost precisely observer-frame $i$), as too were the values of $k$ and $s$.  The best-fitting values are presented in Table \ref{table:GRB_SN_obs_props}.

Using the values of $k$ and $s$ determined from the LC decomposition method, we then performed an identical bootstrap analysis as that in C14 to determine $m$ and $b$ of a line fitted to the combined $k$ and $s$ values for the sample of GRB-SNe from C14, and including those of the SN accompanying GRB~140606B.  Our results are displayed in Fig. \ref{fig:ks}, where find:  $m=1.58\pm0.20$, $b=-0.30\pm0.14$, $r=0.935$ and $p=1.5\times10^{-9}$.  In C14, the slope and $y$-intercept were found to be $m=1.60\pm0.20$ and $b=-0.31\pm0.15$, respectively, and $r=0.936$ and $p=4.2\times10^{-9}$.  It thus appears that the SN associated with GRB~140606B also follows the $k$--$s$ relation first seen in C14, and indeed reinforces the statistical significance of the analogous luminosity--decline relationship.

We also modelled the peak redshifted $V$-band SN properties of the associated SN in using the method presented in Cano \& Jakobsson (2014; CJ14 hereafter).  This time, instead of fitting the observations of the AG \& SN, we removed the AG component and modelled just the SN in order to determine the time of peak light, the peak magnitude, and the $\Delta m_{15}$ parameter (i.e. how much the LC fades in magnitudes from peak light to 15 days later).  We found: $m_{\rm p} = 22.08\pm0.28$, $t_{\rm p} = 17.67\pm3.11$ and $\Delta m_{15} = 0.76\pm0.18$.  The errors include the additional uncertainties associated with the poorly constrained rest-frame extinction, as well as the uncertainty in the temporal decay constant.  All of these parameters can be found in Table \ref{table:GRB_SN_obs_props}.

Finally, assuming a distance modulus of $\mu=41.66$ mag (Table \ref{table:GRB_vitals}), the peak absolute $V$-band magnitude of the SN associated with GRB~140606B is $M_{V, \rm p} = -19.58\pm0.43$, which is only 0.1 mag brighter than SN~1998bw (Galama et al. 1998; Cano et al. 2011a).  In relation to the average absolute magnitudes of the general GRB-SN population, Richardson (2009) found $<M_{V, \rm p}> = -19.2\pm0.2$, with a standard deviation of $\sigma=0.7$ mag.  It appears that the SN associated with GRB~140606B falls at the brighter end of this distribution, though given the size of the error-bars, this result is not terribly significant.

\subsection{Rest-frame Extinction}
\label{sec:rest_frame_extinction}

In a similar analysis to that performed in previous works (Cano et al. 2011a; Cano et al. 2014; Schulze et al. 2014), we attempted to constrain the rest-frame extinction by modelling an optical to X-ray SED constructed of contemporaneous optical, NIR and X-ray data, and fitted it with an absorbed power-law.  The XRT data were extracted using $\emph{Swift}$ tools\footnote{\url{http://heasarc.nasa.gov/lheasoft/}}.   In this analysis we adopted the extinction curve profiles of the SMC, LMC and MW as determined by Pei (1992).  The fit was performed using ISIS (Houck \& Denicola 2000) following the method of Starling et al. (2007), with two absorbers, one being the (fixed) Galactic column ($1.26\times 10^{21}$~cm$^{-2}$; Willingale et al. 2013), and the other being intrinsic to the host galaxy. The abundances used are from Wilms et al. (2000). An SMC dust-extinction model was assumed for the host, while the reddening from the Galaxy was fixed to $E(B-V) = 0.1022$ (Schlafly \& Finkbeiner 2011).

Inspection of the \emph{Swift}-XRT LC reveals that it is poorly sampled\footnote{\url{http://www.swift.ac.uk/xrt_curves/00020384/}}.  However, we downloaded the available data, omitting those taken during the XRT anomaly\footnote{\url{http://www.swift.ac.uk/support/anomaly.php}} and constructed a time-averaged X-ray spectrum from the remaining data.  The time-range of the XRT spectrum was 5.143--5.293 days, with a meantime of 5.214 days.  The resulting background-subtracted spectrum (i.e. the source spectrum; not shown) consisted of only 19 photons.

The optical data (Section \ref{sec:LC_decomposition}) were decomposed to isolate just the AG component in filters $griz$ (i.e. the host and SN contributions were mathematically subtracted).  Due to lack of optical data at the precise time of the X-ray spectrum, the optical data were interpolated to $t-t_{0}=5.214$ days using the fitted model (e.g. Fig. \ref{fig:LC_decomp}).  The SN model from C13 failed to reproduce the $K$-band LC of SN~1998bw at $z=0.384$ at times less than 30 days.  Moreover, host observations in $K$ were not obtained.  We were thus unable to account for these two additional components when decomposing the $K$-band observations. However, at +5.214 days the AG is expected to be considerably brighter than the underlying host, while the SN contribution can be estimated to be roughly 10--15\% of the total emitted light in this filter (Cano et al. 2011b).  With these caveats in mind, the $K$-band observations were simply fit with a SPL and interpolated to +5.214 days.  The resulting optical/NIR data, which were well described by a SPL with a spectral index of $\beta=1.39\pm0.02$ ($\chi^{2}/dof=3.6/3$), were combined with the X-ray spectrum and then fit with a series of PLs and extinction curves.

Unsurprisingly the rest-frame extinction was not very well constrained.  When all data were considered, a value of $E(B-V)_{\rm rest} = 0.16\pm0.14$ mag was obtained at the 90\% confidence level, which is the value adopted in this paper.  We were unable to constrain the rest-frame hydrogen column density in all fits, and to reduce the number of free parameters we simply fixed the entire column density to that of the MW along the sight-line towards GRB 140606B.  The adopted extinction profile is that of the SMC, however fits using the LMC and MW give similar results.

\subsection{Bolometrics}
\label{sec:bolometrics}

\begin{figure}
 \centering
 \includegraphics[bb=0 0 603 403, scale=0.75]{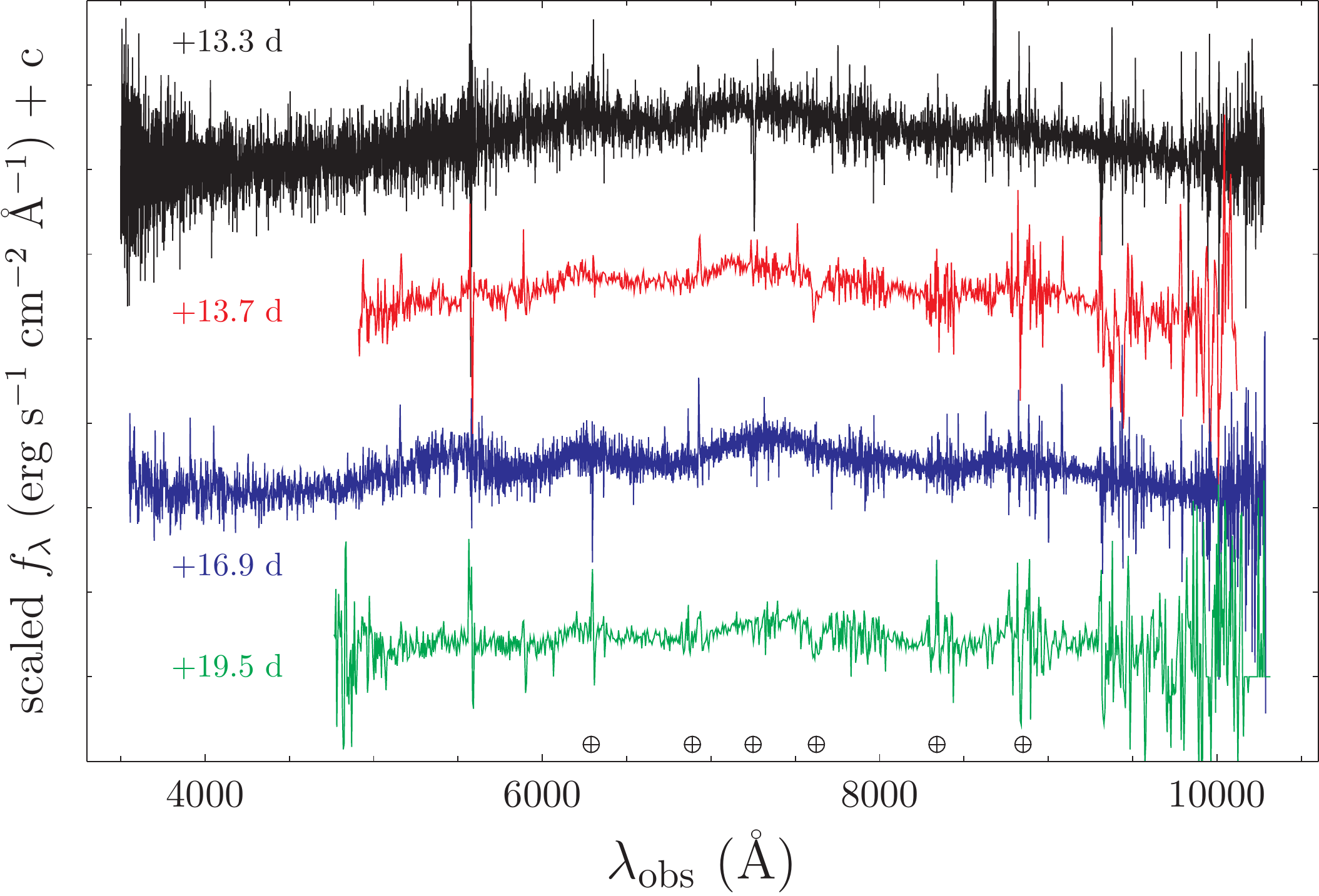}
 \caption{Spectroscopic evolution of the SN associated with GRB 140606B / iPTF14bfu.  Spectra are presented in observer-frame wavelengths, while the epochs are in rest-frame time relative to the time of explosion.  Prominent O$_{2}$ and H$_{2}$O telluric features are indicated with an $\oplus$.  Undulations reminiscent of other SNe Ic-BL, including GRB-SNe, are evident in each epoch of spectroscopy at 6200, 7400, and 8800 \AA.}
 \label{fig:spectra_timeseries}
\end{figure}

\begin{figure}
 \centering
 \includegraphics[bb=0 0 611 403, scale=0.75]{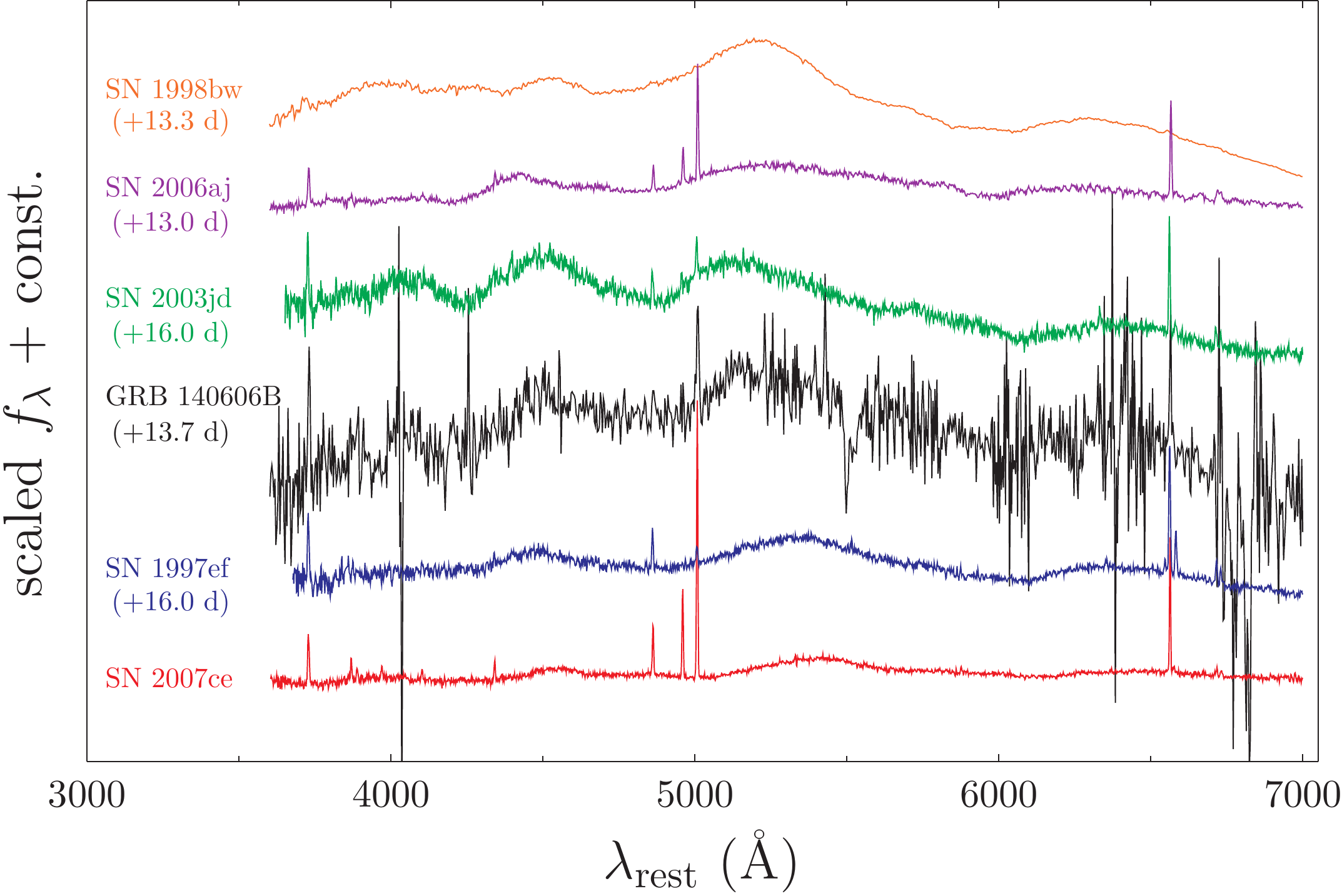}
 \caption{Comparison of the GTC spectrum taken on 25-June-2014 ($t~-~t_{0}~=~13.7$~days; less than a day before peak $i$-band light) compared with other SNe Ic-BL, see main text for appropriate references.  All times and wavelengths are given as rest-frame quantities, and times are relative to their explosion dates.  Note that the explosion date, and time of peak light, are unknown for SN~2007ce.  }
 \label{fig:spectra_compare}
\end{figure}

\begin{table}
\centering
\setlength{\tabcolsep}{5.0pt}
\setlength{\extrarowheight}{3pt}
\caption{GRB 140606B / iPTF14bfu: SN line velocities}
\label{table:SN_line_velocities}
\begin{tabular}{ccc}
\hline					
Transition	&	$t-t_{0}$ (days)	&	$v$ (km s$^{-1}$)	\\
\hline					
Fe \textsc{ii} $\lambda$5169	&	13.30	&	$-20,630\pm1050$	\\
Fe \textsc{ii} $\lambda$5169	&	13.72	&	$-19,820\pm1280$	\\
Fe \textsc{ii} $\lambda$5169	&	16.92	&	$-17,160\pm1430$	\\
Fe \textsc{ii} $\lambda$5169	&	19.46	&	$-15,570\pm1420$	\\
Si \textsc{ii} $\lambda$6355	&	13.30	&	$-16,740\pm3220$	\\
Si \textsc{ii} $\lambda$6355	&	16.92	&	$-13,820\pm2650$	\\
\hline					
\end{tabular}
\begin{flushleft}
NB: Times are given as rest-frame. \\
\end{flushleft}
\end{table}


We calculated the bolometric properties (nickel mass, ejecta mass and kinetic energy, $M_{\rm Ni}$, $M_{\rm ej}$ and $E_{\rm K}$, respectively) of the SN accompanying GRB~140606B / iPTF14bfu for the rest-frame filter range $UBVRIJH$ using the method in C13.  We did not include the uncertain UV contribution to the bolometric LC of SN~1998bw as done in other analyses (Cano et al. 2011b; Schulze et al. 2014; Lyman et al. 2014).  We used the average stretch and luminosity factors of the SN in observer-frame filters $r$ and $i$, neglecting those determined in filters $g$ and $z$ due to the assumptions made in Section \ref{sec:LC_decomposition}, where we had to fix the value of decay index ($g$ and $z$) and the stretch factor ($g$) during the fit.  However, in filters $r$ and $i$, there were enough datapoints in the combined datasets (both ours and those in Singer et al. 2015) to allow all parameters to vary freely during the fit.  The average stretch and luminosity factors were $s_{\rm ave} = 0.83\pm0.10$ and $k_{\rm ave} = 1.19\pm0.31$.  We then modified the $UBVRIJH$ bolometric LC of SN~1998bw by these average values, and then fit the resultant bolometric LC with the Arnett model (Arnett 1982; Valenti et al. 2008).  To determine the ejecta mass and kinetic energy, we used a peak photospheric velocity of $v_{\rm ph} = 19,820 \pm 1280$ km s$^{-1}$, as determined from blueshifted Fe \textsc{ii} $\lambda$5169 in the peak optical spectra (Section \ref{sec:SN_spectra}).  Motivation for using Fe \textsc{ii} $\lambda$5169 as opposed to Si \textsc{ii} $\lambda$6355 comes from analyses presented in the literature, such as Hamuy \& Pinto (2002), Valenti et al. (2009) and Schulze et al. (2014), who show that the former is a better proxy for the photospheric velocity.  Finally, a thorough discussion of the caveats of fitting the Arnett model to bolometric LCs of GRB-SNe is presented in C13.

Using our method we found: $M_{\rm Ni} = 0.42\pm0.17$ $\rm M_{\odot}$, $M_{\rm ej} = 4.8\pm 1.9 $ $\rm M_{\odot}$ and $E_{\rm K} = $ ($1.9\pm 1.1$) $\times 10^{52}$ erg.  The quoted errors include the uncertainties in $v_{\rm ph}$, $k$ and $s$.  These bolometric values are quite typical of the general GRB-SN population: C13 found for the aforementioned sample of $N=20$ GRB-SNe a median nickel mass of $\tilde{M}_{\rm Ni} = 0.34$~M$_{\odot}$  ($\sigma=0.24$~M$_{\odot}$), a median ejecta mass of $\tilde{M}_{\rm ej} = 5.9$~M$_{\odot}$ ($\sigma=3.9$~M$_{\odot}$), and a median kinetic energy of $\tilde{E}_{\rm K} = 2.2 \times 10^{52}$ erg ($\sigma=1.5 \times 10^{52}$~erg).  The ratio $\frac{M_{\rm Ni}}{M_{\rm ej}} = 0.09$ is also consistent with that measured for the sample of GRB-SNe, where an average value of 0.07 ($\sigma=0.04$) was determined in C13.  Moreover, the peak photospheric velocity is very similar to the average peak photospheric velocities measured for a sample of GRB-SNe by C13 ($\bar{v}_{\rm ph}= 20,000\pm2500$ km s$^{-1}$) and for a sample of SNe IcBL (which included non-GRB-SNe) by Lyman et al. 2014 ($\bar{v}_{\rm ph} =19,100\pm5000$ km s$^{-1}$).

\section{Spectroscopic Analysis}
\label{sec:SN_spectra}

\begin{figure}
 \centering
 \includegraphics[bb=0 0 576 432, scale=0.8]{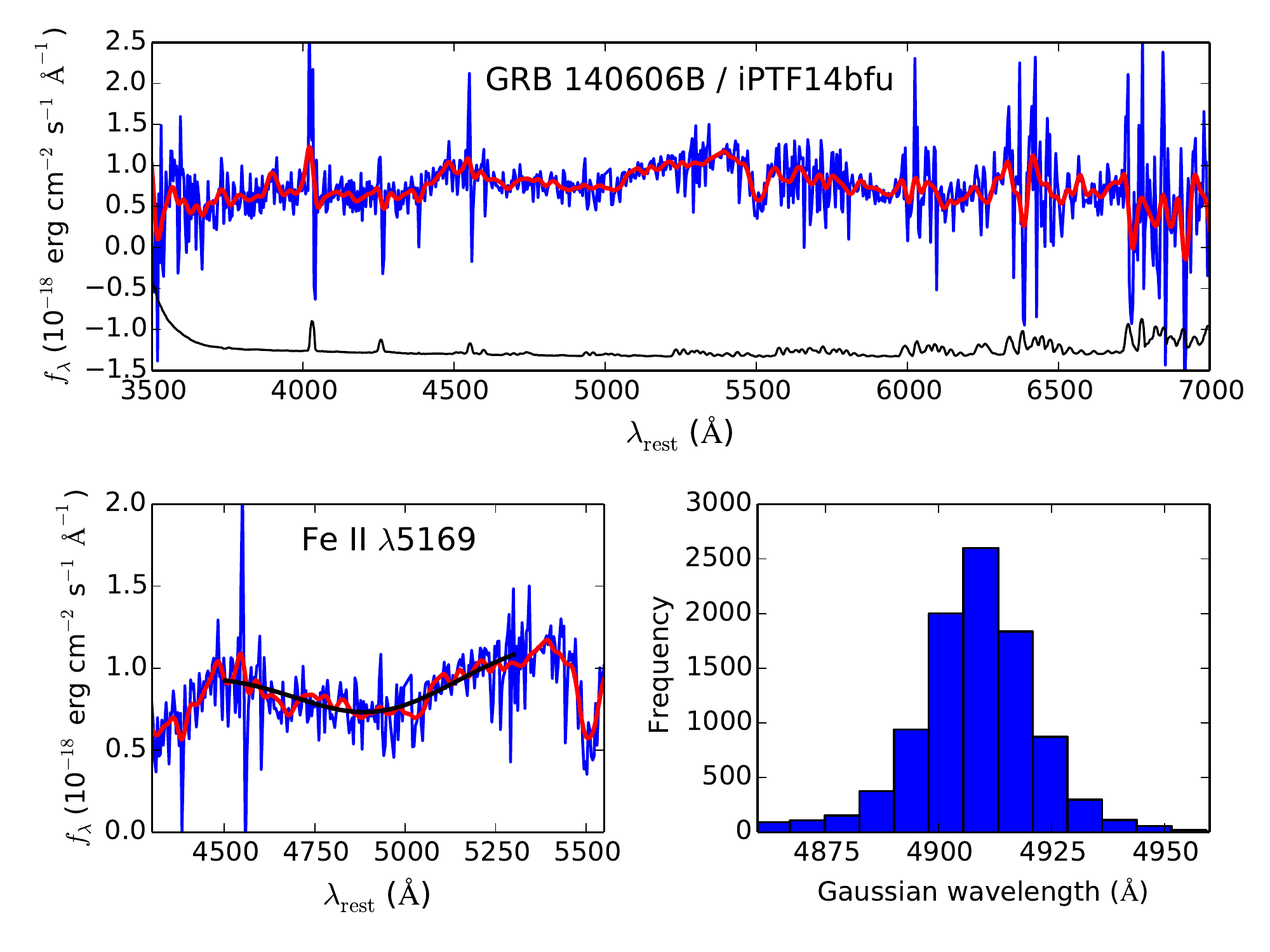}
 \caption{An example of our line-fitting procedure on the GTC spectrum from 03-July-2014 (+19.5 days, rest-frame) to determine the central wavelength and blueshifted velocity of the absorption feature seen between 4500--5400 \AA, which we attribute to blueshifted Fe \textsc{ii} $\lambda$5169.  The original GTC spectrum is shown in blue, the smoothed spectrum is in red, and the error spectrum in black, which has been arbitrarily shifted for visual purposes.  A MC simulation was performed that fit a single Gaussian to the feature, which was repeated 10,000 times, where in each step a new spectrum was created from the original by using random sampling from the error spectrum, while the smoothing parameters ($M$ and $\beta$) were also allowed to randomly vary between sensibly chosen limits.  In this example we determined a central wavelength of $\lambda=4907\pm23$ \AA, which corresponds to a blueshifted velocity of $v=-15,570\pm1420$ km s$^{-1}$.}
 \label{fig:spectral_modelling}
\end{figure}

\begin{figure}
 \centering
 \includegraphics[bb=0 0 281 181, scale=1.5]{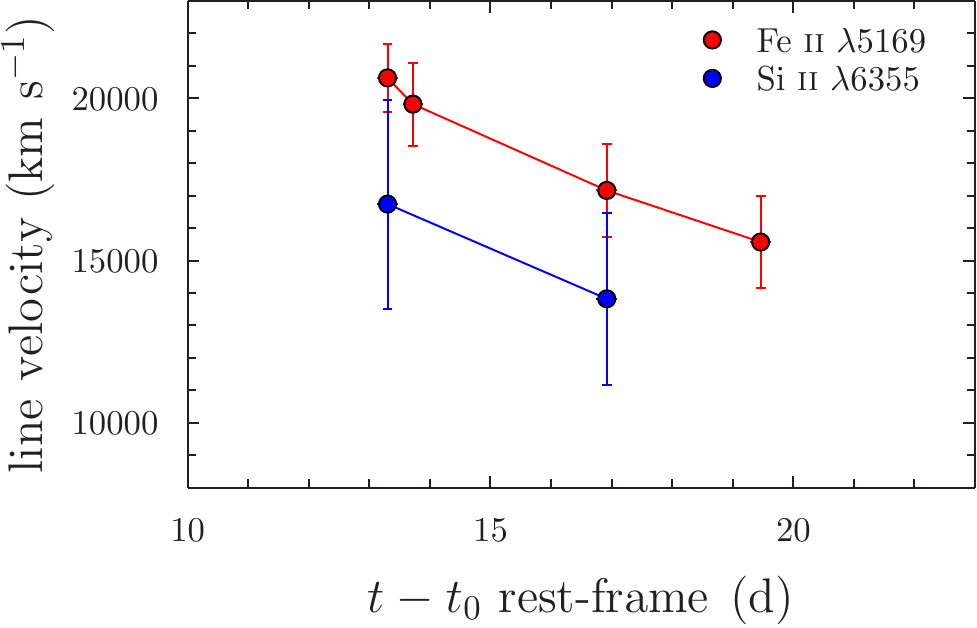}
 \caption{Temporal evolution of the blueshifted line velocities of Si \textsc{ii} $\lambda$6355 (blue) and Fe \textsc{ii} $\lambda$5169 (red).  }
 \label{fig:SN_line_velocities}
\end{figure}

Of the several epochs of spectroscopy obtained with the GTC and Keck, four epochs show clear SN features.  The time-series spectra of the SN associated with GRB~140606B is presented in Fig. \ref{fig:spectra_timeseries}, which span a time range of $t-t_{0}=13.3-19.5$ days in the rest-frame.  Peak, rest-frame, $V$-band light occurred at $\approx 14.0$ days rest-frame, meaning we obtained spectra of the SN just before and after peak light.  Strong telluric lines (O$_{2}$ and H$_{2}$O) are indicated with an $\oplus$, and the spectra are not corrected for extinction.  All spectra display undulations typical of SN spectra, including clear bumps at (observer-frame) 6200, 7400, and 8800 \AA.

In Fig. \ref{fig:spectra_compare} we compared the GTC spectrum at $t-t_{0}=13.7$~days with other SNe Ic-BL: GRB~980425 / SN~1998bw (Patat et al. 2001), GRB~060218 / SN~2006aj (Pian et al. 2006), GRB~100316D / SN~2010bh (Bufano et al. 2012), SN~2003jd (Valenti et al. 2008; Modjaz et al. 2014), SN~1997ef (Modjaz et al. 2014) and SN~2007ce (Modjaz et al. 2014).  The GRB-SNe epochs were chosen so to compare the spectra of GRB-SNe at similar times from the time of explosion, while the other SNe Ic-BL not associated with GRBs were chosen to provide a suitable comparison of the velocities of key absorption features, especially Si \textsc{ii} $\lambda$6355, with that of the SN accompanying GRB~140606B.

Despite the modest S/N of the spectra in Fig. \ref{fig:spectra_timeseries} we attempted to model two key absorption features: Si \textsc{ii} $\lambda$6355 and Fe \textsc{ii} $\lambda$5169.  Si \textsc{ii} is only measurable in the Keck spectra due to a slightly better telluric correction than the GTC spectra, however the broad Fe \textsc{ii} feature was measurable in all four epochs.  Using a similar analysis to that performed in Cano et al. (2014), we used a program written in \textsc{python} to first smooth each spectrum and then fit a single Gaussian to the smoothed feature of interest in order to determine its central wavelength and blueshifted velocity.  The spectrum was smoothed using a Kaiser window as a smoothing kernel, where the size of the window ($M$) and shape of the window ($\beta$) were varied.  In order to get an estimate of the error of each wavelength (and velocity) measurement, we performed a bootstrap analysis with Monte Carlo (MC) sampling to generate 10,000 spectra from the original spectrum and its error spectrum.  At each wavelength in the original spectrum we derived a random number from a Gaussian distribution that is centered at each wavelength in the spectrum, and whose standard deviation is equal to the value of the error spectrum at that wavelength.  We also allowed $M$ and $\beta$ to be randomly chosen between two pre-determined values in the simulation.  

The result of such a simulation to the GTC spectrum obtained on 03-July-2014 is shown in Fig. \ref{fig:spectral_modelling}.  In this example we determined the central wavelength of the absorption feature between 4500--5400 \AA, which we attribute to blueshifted Fe \textsc{ii} $\lambda$5169, to be $\lambda=4970\pm23.4$ \AA, which corresponds to a blueshifted velocity of $v=-15,570\pm1420$ km s$^{-1}$.   The GTC spectrum obtained on 25-June-2014 corresponds to the closest date to peak $V$-band light, where we measured a velocity of $v=-19,820\pm1280$ km s$^{-1}$, which we use a close proxy to the peak photospheric velocity when determining the bolometric properties of the associated SN in Section \ref{sec:bolometrics}.  Identical simulations were performed on all four spectra, where we also determined the central Gaussian wavelengths and corresponding blueshifted velocities of Si \textsc{ii} $\lambda$6355 in the two Keck spectra.  Note that we also attempted to fit double Gaussians to each profile, but no improvement was obtained in the fits.  Our results are presented in Table \ref{table:SN_line_velocities}, while the temporal evolution of the line velocities are presented in Fig. \ref{fig:SN_line_velocities}.

As noted above, the blueshifted Si \textsc{ii} $\lambda$6355 in our spectra unfortunately occurs at the same wavelength range as a series of telluric lines (in the region 9000--10,000 \AA), which hindered our abilities to fit it with Gaussian profiles to determine its blueshifted velocity (e.g. Cano et al. 2014), especially in the GTC spectra.  However, comparison of the rest-frame GTC spectrum from 25-June-2015 (Fig. \ref{fig:spectra_compare}) with those of SN~2003jd and SN~1997ef show that the absorption feature near (rest-frame) 6100 \AA $ $ coincide well together.  Indeed when this GTC spectrum was compared to a series of SN templates using SNID version 5.0 (Blondin \& Tonry 2007), which includes the templates of Lui \& Modjaz (2014), good matches were obtained to SN~2003jd, and SN~1997ef from -4 days to +5 days from peak light.  Prompted by this initial analysis, it was found that good fits were obtained to the spectra of SN~2003jd taken 01-Nov-2003, which is estimated to be +3 days from peak $B$-band light, and $\approx +16$ days from the date of explosion.  Valenti et al. (2008) determined, for this epoch, a blueshifted velocity of Si \textsc{ii} $\lambda$6355 of $v=13,500\pm1000$ km s$^{-1}$.  Additionally, the spectrum of SN 1997ef obtained on 06-Dec-1997 (+6 days from peak B-band light; Mazzali et al. 2000) proved also to be a good fit to the spectrum of GRB 140606B. Mazzali et al. (2000) estimated from their spectral synthesis models a blueshifted velocity of Si \textsc{ii} $\lambda$6355 of  $\approx 13,000$ km s$^{-1}$.  In comparison, the blueshifted velocity of Si \textsc{ii} $\lambda$6355 at the time of the GTC epoch near peak light is roughly $16,000\pm3500$ km s$^{-1}$ (Fig. \ref{fig:SN_line_velocities}).  The line velocities of all the SNe are consistent within their respective error-bars.

\section{Host Galaxy and Companion}
\label{sec:host_spectra}

\begin{table}
\centering
\setlength{\tabcolsep}{5.0pt}
\setlength{\extrarowheight}{3pt}
\caption{GRB 140606B / iPTF14bfu: GTC host photometry observation log}
\label{table:photometry_HOST_obs_log}
\begin{tabular}{cccc}
\hline																														
Object	&	$t-t_{0}$ (days)	&	Filter	&		mag$^{a}$				\\
\hline
Host	&	171.774	&	$g$	&	$	25.59	\pm	0.12	$	\\
Host	&	171.784	&	$r$	&	$	24.50	\pm	0.07	$	\\
Host	&	171.794	&	$i$	&	$	24.30	\pm	0.09	$	\\
Host	&	171.805	&	$z$	&	$	23.98	\pm	0.12	$	\\
\ldots	&	\ldots	&	\ldots	&	\ldots					\\
Companion	&	171.774	&	$g$	&	$	24.96	\pm	0.08	$	\\
Companion	&	171.784	&	$r$	&	$	24.26	\pm	0.06	$	\\
Companion	&	171.794	&	$i$	&	$	23.92	\pm	0.07	$	\\
Companion	&	171.805	&	$z$	&	$	23.91	\pm	0.12	$	\\
\hline
\end{tabular}
\begin{flushleft}
$^{a}$ Apparent magnitudes, in the AB system, are not corrected for foreground or rest-frame extinction. \\
\end{flushleft}
\end{table}


\begin{figure}
 \centering
 \includegraphics[bb=0 0 1122 828, scale=0.4]{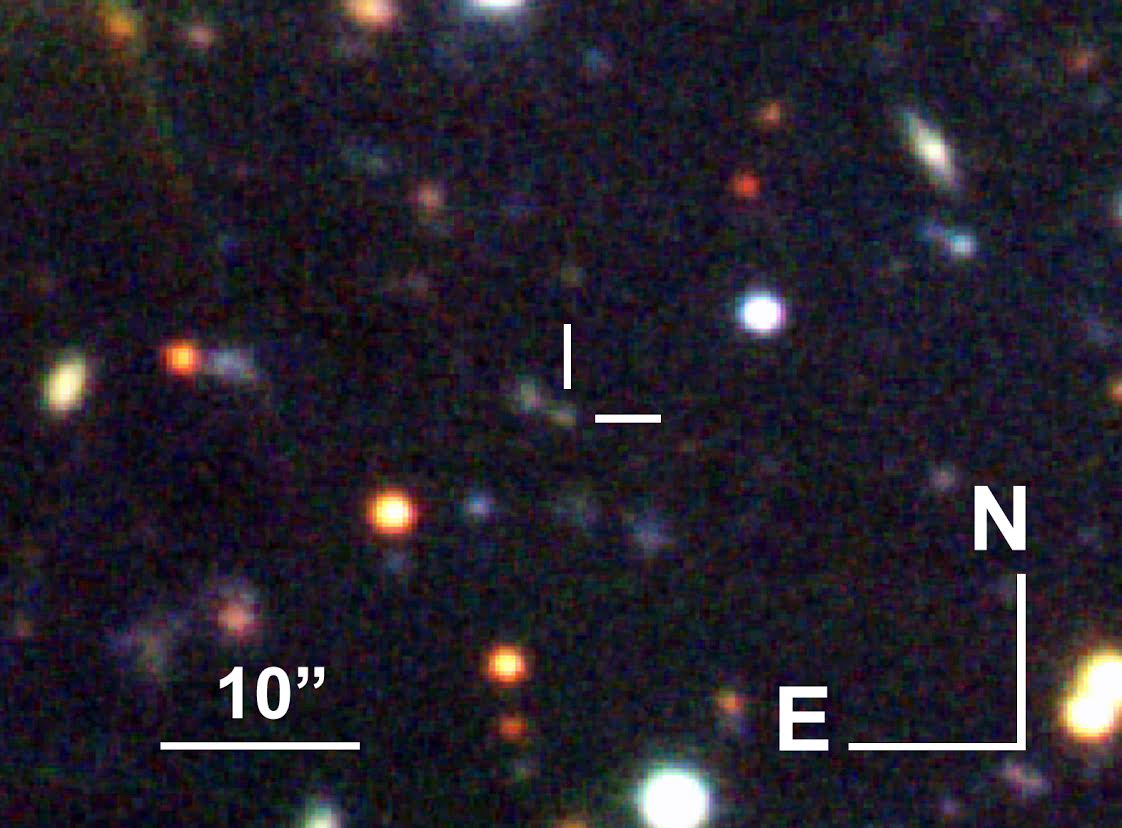}
 \caption{RGB image of GRB 140606B constructed from late-time GTC images taken in filters $g$ (blue), $r$ (green) and $i$ (red).  The position of the host galaxy is indicated by the black crosshairs.  A companion galaxy at the same redshift as the GRB host ($z=0.384$) is seen just to the left, where the projected distance between them is $\approx$4.7 kpc. }
 \label{fig:Host_finder}
\end{figure}

In the late-time GTC images, the host galaxy and the nearby companion galaxy are clearly detected (see Fig. \ref{fig:Host_finder}).  The average separation of a circular aperture centered on each galaxy (as determined when using IRAF digiphot.daophot.centroid) in the $griz$ images is $\approx0.88''$, which corresponds to a projected distance of $\approx4.7$ kpc.

\begin{figure}
 \centering
 \includegraphics[bb=0 0 668 820, scale=0.65]{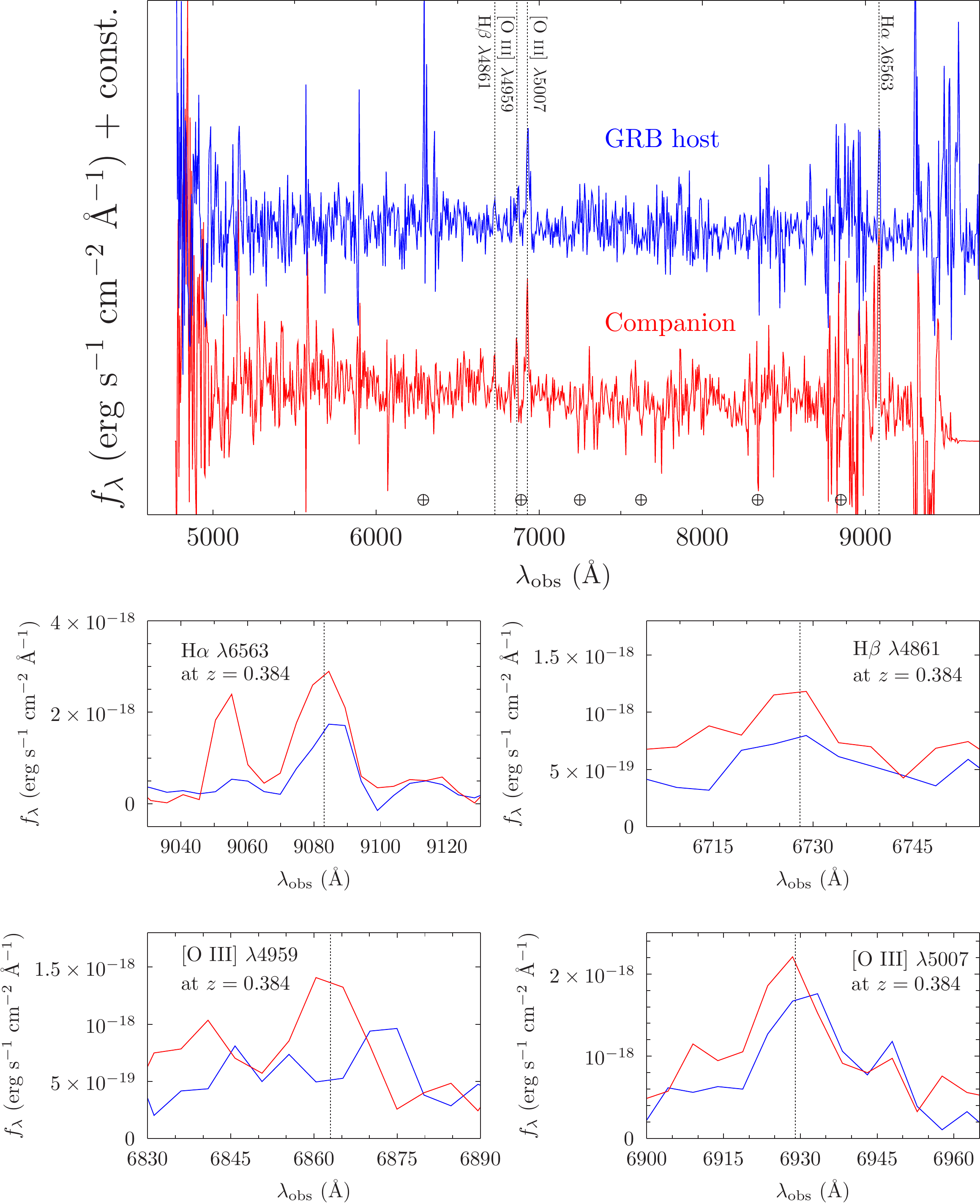}
 \caption{Comparison of the GTC spectra of the GRB host (blue) and the nearby companion galaxy (red).  Both spectra show prominent H$\alpha$ and [O \textsc{iii}] $\lambda5007$ emission lines, and weak H$\beta$ emission lines.  The companion also shows a redshifted [O \textsc{iii}] $\lambda4959$ emission line.  The line fluxes and redshifts of the various lines are presented in Table \ref{table:host_lines}, and the derived Balmer decrements and SFRs in Table \ref{table:host_SFR}.  Note that in the top figure, the spectrum of the GRB host galaxy has been arbitrarily shifted for presentation purposes, however, in the lower panels the dereddened and flux-calibrated spectra are presented with no further alterations.  Thus the relative fluxes seen between the two galaxies are real.}
 \label{fig:host_spectra}
\end{figure}

\begin{table}
\centering
\setlength{\tabcolsep}{2.0pt}
\setlength{\extrarowheight}{3pt}
\caption{Absorption \& Emission lines of GRB Host and Companion (Other)}
\label{table:host_lines}
\begin{tabular}{cccccc}
\hline											
Galaxy	&		&	$\lambda_{\rm rest}$ (\AA)	&	$\lambda_{\rm obs}$ (\AA)	&	$z$	&	$f^{\ddagger}$	\\
\hline											
GRB	&	H$\alpha$	&	6562.80	&	$9085.09\pm0.67$	&	$0.3843\pm0.0001$	&	$2.17\pm0.21$	\\
GRB	&	H$\beta$	&	4861.33	&	$6728.90\pm1.50$	&	$0.3842\pm0.0002$	&	$0.72\pm0.23$	\\
GRB	&	[O \textsc{iii}]	&	5006.84	&	$6931.36\pm0.91$	&	$0.3843\pm0.0002$	&	$2.52\pm0.30$	\\
\ldots	&	\ldots	&	\ldots	&	\ldots	&	\ldots	&	\ldots	\\
Other	&	H$\alpha$	&	6562.80	&	$9082.70\pm0.51$	&	$0.3840\pm0.0001$	&	$3.19\pm0.36$	\\
Other	&	H$\beta$	&	4861.33	&	$6725.71\pm2.63$	&	$0.3835\pm0.0005$	&	$0.67\pm0.36$	\\
Other	&	[O \textsc{iii}]	&	4958.92	&	$6862.39\pm1.44$	&	$0.3838\pm0.0003$	&	$0.68\pm0.19$	\\
Other	&	[O \textsc{iii}]	&	5006.84	&	$6927.45\pm0.84$	&	$0.3836\pm0.0002$	&	$1.75\pm0.30$	\\
\hline											
\end{tabular}
\begin{flushleft}
$^{\ddagger}$ units of $10^{-17}$ erg s$^{-1}$ cm$^{-2}$. \\
NB: Wavelengths are in air, and the spectra have been corrected for foreground extinction. \\
\end{flushleft}
\end{table}


The GTC spectra of the host galaxy and the nearby companion were dereddened (foreground only) using IRAF (noao.onedspec.deredden) and flux calibrated using the contemporaneous GTC photometry.  Prominent emission lines, namely Balmer $\alpha$ and $\beta$ lines, and singly and doubly ionised oxygen, were then modelled using \textsc{python}.  The fitting process consisted of a general bootstrap analysis, where single and double Gaussians were fit to the emission lines in order to determine the central wavelength of the fitted Gaussian(s), and the integrated line flux.  In the bootstrap algorithm, MC sampling was performed, where the flux value at a given wavelength was determined using the error spectra, which defined the limits that the sampling occurred within.  10,000 simulations were performed each time, and the standard deviation of the fitted wavelength and the calculated line fluxes were taken as the standard error of each quantity.  These were then added in quadrature to the uncertainty in the GTC wavelength calibration, which was taken to be 0.14 \AA.  The best-fitting values for the different emission lines, which are in air, are displayed in Table \ref{table:host_lines}.


The presence of H$\alpha$ and H$\beta$ allow us to calculate the Balmer decrement for each galaxy, as well as estimate the star-formation rate (SFR) of each.  For the Balmer decrement, we use the following equation (e.g. Reynolds et al. 1997):

\begin{equation}
 E(B-V) = 2.21 \times \mathrm{log_{10}\frac{H{\alpha}/H{\beta}}{2.76}}
\end{equation} 

\noindent where we assume an intrinsic ratio of H${\alpha}$/H${\beta}$=2.76.

Next, we used the following equations from Savaglio et al. (2009) for the SFR derived using H${\alpha}$, which differ to that presented in Kennicutt (1998) as the former authors use the more realistic initial mass function proposed by Baldry \& Glazebrook (2003):

\begin{equation}
 \mathrm{SFR(H{\alpha})} = 4.39 \times 10^{-42} ~\mathrm{L(H{\alpha}) ~~M_{\odot}~yr^{-1}}
\end{equation}


\noindent We used the luminosity distance presented in Table \ref{table:GRB_vitals} to convert the emission line fluxes into luminosities, where the latter are in units of erg s$^{-1}$.

The calculated (dereddened) Balmer decrements and SFRs of both galaxies are presented in Table \ref{table:host_SFR}.  The calculated SFRs of each galaxy are quite similar, and are in the range 0.05--0.08 M$_{\odot}$ yr$^{-1}$.  The integrated Balmer decrement for the GRB's host implies a lower amount of extinction ($E(B-V)\approx 0.1$ mag) than the companion galaxy ($E(B-V)\approx 0.5$ mag).  However, the poorly constrained line flux for H$\beta$ in both spectra (see Fig. \ref{fig:host_spectra}), especially that of the companion galaxy, means we are not able to determine the global extinction value for either very accurately.  However, the value of $E(B-V)$ for the entire host of GRB~140606B is consistent with the poorly constrained value of the extinction local to the GRB itself (Section \ref{sec:rest_frame_extinction}), though of course they do not necessarily need to be, nor usually are, the same value. 

The average redshift of both galaxies is $z= 0.384$.  More specifically, the redshift of the GRB host galaxy, taking the average of the redshift determined from the Balmer and [O \textsc{ii}] emission lines is $0.3843\pm0.0001$.  Similarly, the average redshift of the companion galaxy determined from the emission lines is $0.3837\pm0.0005$.  


\begin{table}
\centering
\setlength{\tabcolsep}{2.5pt}
\setlength{\extrarowheight}{3pt}
\caption{Physical Properties of GRB Host and Companion}
\label{table:host_SFR}
\begin{tabular}{ccc}
\hline							
Galaxy	&	SFR(H$\alpha$) (M$_{\odot}$ yr$^{-1}$)	&	$E(B-V)$ (mag)$^{\dagger}$	\\
\hline					
GRB	&	$0.052\pm0.005$	&	$0.09^{+0.27}_{-0.09}$	\\
Companion	&	$0.077\pm0.006$	&	$0.52^{+0.67}_{-0.35}$	\\
\hline							
\end{tabular}
\begin{flushleft}
$^{\dagger}$ Balmer decrement. \\ 
NB: These are the unobscured SFRs as determined from the dereddened spectra.  The presented errors are statistical, as determined from the modelling procedure. \\
\end{flushleft}
\end{table}

\section{Is GRB 140606B a SBO-GRB or a jetted-GRB?}
\label{sec:SBO_vs_jet}

\subsection{The $E_{\rm p}$--$E_{\rm iso, \gamma}$ plane}

\begin{figure}
 \centering
 \includegraphics[bb=0 0 265 172, scale=1.7]{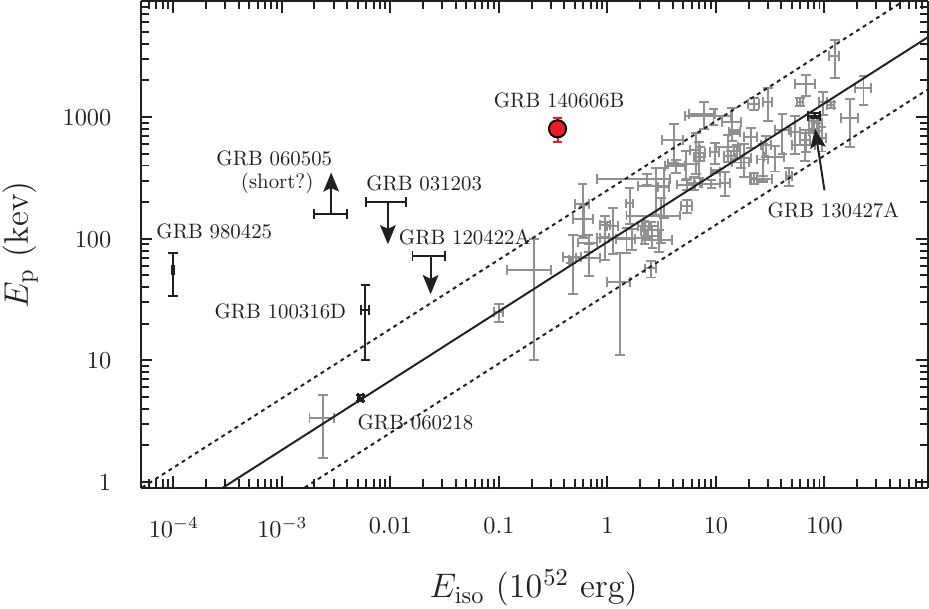}
 \caption{GRB 140606B (red) and the Amati relation ($E_{\rm p}$ versus $E_{\rm iso, \gamma}$).  Data from Amati (2002; 2007; 2008) are shown in grey along with their best fit to a single power-law ($\alpha=0.57$) and the 2-$\sigma$ uncertainty in their fit.  Also noted by Singer et al. (2015), GRB 140606B is an outlier in the Amati relation, and occupies the same region as other $ll$GRBs (shown in black) such as GRBs 980425, 031203 and 100316D, but not GRB 060218, which is fully consistent with the relation.  }
  \label{figure:amati}
\end{figure}

Singer et al. (2015) asked the question of whether the $\gamma$-ray emission detected for GRB~140606B arose from prompt emission generated by a relativistic jet (i.e. a jetted-GRB), as expected for most GRBs, or whether it arose from high-energy emission arising from a shock breakout (SBO), such as attributed to low-luminosity GRBs ($ll$GRBs) 060218 (Campana et al. 2006; though see Friis \& Watson 2013) and 100316D (Starling et al. 2011; Cano et al. 2011b).  Their motivation for this hypothesis arose from the fact that GRB~140606B is an outlier in the Amati relation, and occupies the same region as $ll$GRBs 980425, 031203, 100316D, the intermediate-GRB 120422A, and the SN-less GRB 060505 (Fynbo et al. 2006; Ofek et al. 2007)\footnote{The latter has been suggested to be a short burst, but based on its host environment, spectral lag, and duration it  most likely belongs to the long-burst class (Th\"one et al. 2008, McBreen et al. 2008, Bromberg et al. 2013).}  Singer et al. (2015) displayed the position of GRB~140606B in the $E_{\rm p}$--$E_{\rm iso, \gamma}$ plane (i.e. the Amati relation), which we have reproduced here for the sake of this discussion (Fig. \ref{figure:amati}).  Data included in the plot are taken from Amati et al. (2002, 2007, 2008), while data for GRB~100316D are from Starling et al. (2011), data for GRBs 120729A, 130215A and 130831A are from Cano et al. (2014), and data for GRB~120422A are from Schulze et al. (2014) and Melandri et al. (2012).  It is seen that GRB~140606B deviates from the Amati relation by more than 2$\sigma$, and can be considered to be a hard, weak LGRB in the sense that the value of $E_{\rm \gamma, iso}$ is much lower than expected for its value of $E_{\rm p}$.

In this discussion there are several points that should be noted.  Starling et al. (2011) reported that \emph{Swift} was not able to fully observe the entire $\gamma$-ray emission of GRB~100316D.  This implies that $E_{\rm iso, \gamma}$ is likely underestimated for this event -- any additional $\gamma$-ray emission would push this event closer to the expected range of the Amati relation, and hence it would be less of an outlier.  Moreover, $E_{\rm p}$ determined by Fan et al. (2011) is poorly constrained.  Therefore it is possible that GRB~100316D is not an outlier of the Amati relation at all.

Additionally, The gamma-ray properties of GRB~120422A are also not well constrained. Melandri et al. reported a peak energy of $E_{\rm p} = 33^{+39}_{-33}$ keV, with an upper limit of 72 keV at the 90\% confidence limit.  The latter value is the one adopted here (see Fig. \ref{figure:amati}).  Had we adopted the former range, GRB~120422A is actually consistent with the Amati relation at 2$\sigma$.

\subsection{The $M_{V, \rm p}$--$L_{\rm iso, \gamma}$ plane}

\begin{figure}
 \centering
 \includegraphics[bb=0 0 276 177, scale=1.6]{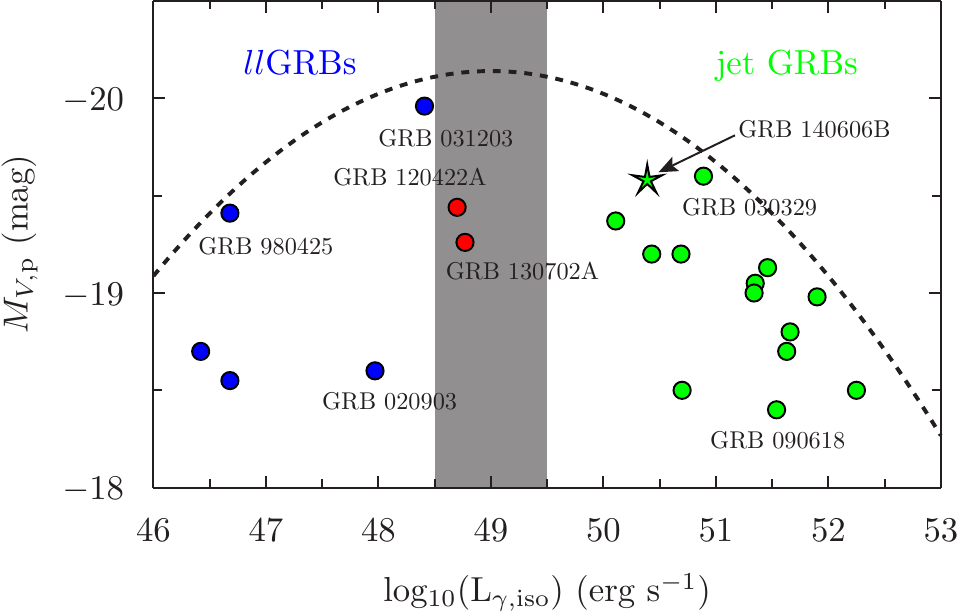}
 \caption{Peak (rest-frame) $M_{V}$ magnitudes of a sample of GRB-SNe.  $ll$GRBs ($L_{\gamma,\rm iso} < 48.5$) are displayed in blue, jetted-GRBs ($L_{\gamma,\rm iso} > 49.5$) are shown in green, and two intermediate GRBs (120422A and 130702A) are presented in the shaded grey area in red.  GRB~140606B (green star) is found in the region occupied by jetted GRBs.  As originally noted by Hjorth (2013), there appears to be a parabola-shaped upper-envelope to the peak $V$-band brightness of a GRB-SN as a function of its $\gamma$-ray luminosity.  See the main text for references to the various GRB-SN datasets. }
 \label{fig:Mv_Liso_compare}
\end{figure}

Hjorth (2013) plotted the peak $V$-band magnitudes of a sample of GRB-SNe against their isotropic-equivalent luminosity in $\gamma$-rays, where the latter is defined as $L_{\gamma,\rm iso} = E_{\gamma,\rm iso}~(1+z)~t_{90}^{-1}$.  We have reproduced that plot here (Fig. \ref{fig:Mv_Liso_compare}) using data from Hjorth \& Bloom (2012) for GRB-SNe of grades A--C, C13, CJ14, and D'Elia et al. (2015).  It is seen that GRB~140606B falls in the jetted-GRB distribution, as opposed to the $ll$GRBs and the two intermediate GRBs 120422A (Schulze et al. 2014) and 130702A (D'Elia et al. 2015), though we note that the intermediate nature of GRB~130702A was not discussed by D'Elia et al. (2015).  Also plotted is a parabola-shaped upper-envelope (aka upper limit) to the peak $V$-band brightness of a GRB-SN as a function of its $\gamma$-ray luminosity, which was first reported by Hjorth (2013).

In the context of Fig. \ref{fig:Mv_Liso_compare}, GRB~140606B is fully consistent with other jetted-GRBs, and is inconsistent with $ll$GRBs and intermediate GRB-SNe.  With a value of log$_{10}(L_{\gamma,\rm iso}) = 50.4$, it is almost one order of magnitude above the upper limit for intermediate GRBs, and two orders of magnitude above that of the $ll$GRBs imposed by Hjorth (2013).  Using \textit{just} this line of argument, it would be more appropriate to consider GRB~140606B as a jetted-GRB.

\subsection{The $E_{\rm K}$--$\Gamma \beta$ plane}

\begin{figure*}
 \centering
 \includegraphics[bb=0 0 566 425, scale=0.85]{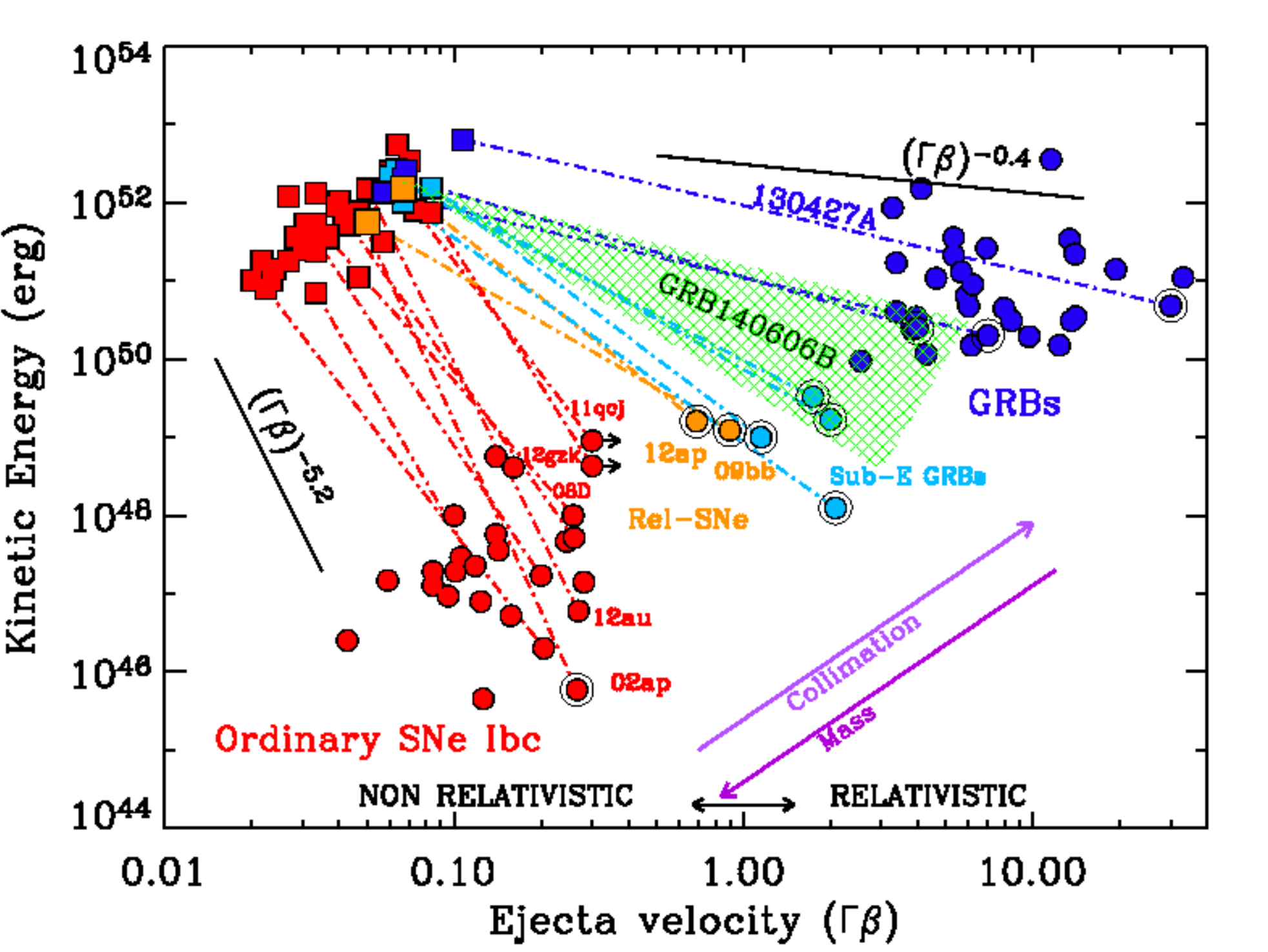}
 \caption{The position of GRB~140606B in the $E_{\rm K}$--$\Gamma \beta$ plane.  Ordinary SNe Ibc are shown in red, $ll$GRBs are shown in light blue,  relativistic SNe IcBL in orange, and jetted-GRBs in dark blue.  Data presented here are from Margutti et al. (2014) and references therein.  Squares and circles are used for the slow-moving and the fast-moving ejecta, respectively, as determined from modelling of optical and radio observations of the respective events.  Open black circles identify explosions with broad lines in their optical spectra.  The velocity of the fast-moving ejecta was computed for $t-t_{0}= 1$ day (rest-frame).   The black solid lines correspond to ejecta kinetic energy profiles of a purely hydrodynamical explosion $E_{\rm K} \propto (\Gamma\beta)^{-5.2}$ (Sakurai 1960; Matzner \& McKee 1999; Tan et al. 2001), explosions that powered by a short-lived central engine (i.e. a SBO-GRB or a relativistic Ic-BL SN such as SNe 2009bb and 2012ap; $E_{\rm K} \propto (\Gamma\beta)^{-2.4}$), and those arising from a long-lived central engine (i.e. a jetted-GRB; $E_{\rm K} \propto (\Gamma\beta)^{-0.4}$; Lazzati et al. 2012).  The high-energy, relativistic properties of GRB~140606B have been determined using its value of $E_{\gamma, \rm iso}$ and estimates for its expected $\gamma$-ray efficiency and an upper-limit to its opening angle as constrained by observations of its optical LC.  Despite our conservative estimates of these properties, it is seen that GRB~140606B occupies an intermediate region between SBO-GRBs and jetted-GRBs, and cannot be unambiguously classified as either one type.}
 \label{fig:EK_GammaBeta_plane}
\end{figure*}

The third line of argument is the placement of the relativistic and non-relativistic ejecta of GRB~140606B in the $E_{\rm K}$--$\Gamma \beta$ plane (e.g. Soderberg et al. 2010; Margutti et al. 2013; Margutti et al. 2014).  Here $\Gamma$ is the bulk Lorentz factor, and $\beta=v/c$.  It was shown by these authors that SNe Ibc arising from different explosion mechanisms have different ejecta kinetic energy profiles.  For SNe that arise from neutrino-driven hydrodynamical explosions, $E_{\rm K} \propto (\Gamma\beta)^{-5.2}$ (Sakurai 1960; Matzner \& McKee 1999; Tan et al. 2001\footnote{Strictly speaking, Tan et al. (2001) showed that there is in fact a range of slopes, and that the slope depends on the ejecta density profile  Their general conclusion is the prediction that the hydrodynamic collapse of a massive star with an ordinary ejecta density profile will have a steep $E_{k}$ profile, one that is steeper than expected for the other two scenarios discussed here.  Conversely, both Sakurai (1960) and Matzner \& McKee (1999) showed that the acceleration of the SBO across the sharp density drop of the envelope near the stellar edge dictates that $E_{\rm K} \propto (\Gamma\beta)^{-5}$. }), whereas explosions that are powered by a short-lived central engine (i.e. a SBO-GRB or a relativistic Ic-BL SN such as SNe 2009bb and 2012ap) have $E_{\rm K} \propto (\Gamma\beta)^{-2.4}$, while those arising from a long-lived central engine (i.e. a jetted-GRB) have $E_{\rm K} \propto (\Gamma\beta)^{-0.4}$ (Lazzati et al. 2012).   

Determining the position of GRB~140606B in this plane can reveal clues as to the origin of its high-energy emission.  In order to do this, the two components of the ejecta need to be considered separately.  In terms of the non-relativistic ejecta, i.e. that arising from the SN itself, we have already determined its kinetic energy to be $E_{\rm K} = 1.9 \times 10^{52}$ erg.  We can then use the peak photospheric velocity of the SN ejecta, $v = 19,820$ km s$^{-1}$ as a proxy for the value of $\Gamma\beta$ as the mass of the SN ejecta is so large there is effectively no deceleration.  We thus obtain a value of $\Gamma\beta=0.066$.

Determining the kinetic energy of the fireball and the ejecta velocity of the high-relativistic ejecta is less straight-forward.  Ideally, the kinetic energy of the fireball should be determined from late-time AG data (Zhang et al. 2007), and preferably radio data (Soderberg et al. 2010; Chakraborti et al. 2014).  Alternatively, it is possible to estimate the kinetic energy from early X-ray data, using e.g. equation 13 from Zhang et al. (2007) for data at $t-t_{0}=1$ day.  However, as noted earlier, the \emph{Swift}/XRT X-ray LC is poorly sampled, and all data obtained before $4\times10^{5}$ s (observer-frame) were taken during the XRT anomaly, meaning these data are far from reliable.  Moreover, the first XRT data-point was obtained at $t-t_{0}=2.15$ day (1.55 day rest-frame), with no prior knowledge of the X-ray LC evolution before this time.  Therefore use of this data to determine the kinetic energy of the fireball at $t-t_{0}=1$ day will be highly uncertain and unreliable.

Another alternative strategy is is to turn to the prompt emission itself and use the known equation for the radiative efficiency of the fireball ejecta ($\eta_{\gamma} = E_{\gamma} / (E_{\rm K} + E_{\gamma})$; where $E_{\gamma} = (\theta^{2}/2)E_{\gamma, \rm iso}$).  With an estimate of the opening angle and the radiative efficiency we can in turn estimate the kinetic energy of the fireball.  Zhang et al. (2007) estimated $\eta$ for a sample of 31 GRBs at two times in the fireball's evolution: at the deceleration time and at the break time.  Their calculated values span a range of $\eta_{\gamma}$=1--99\%, which a rough average in the range $\eta_{\gamma}$=30--60\%.  Based on their result, and in the interest of conservatism, we considered the range $\eta_{\gamma}$=10--90\% in our calculations.

Next, we need to make an estimate of the opening angle of the jet ($\theta$).  From Sari et al. (1999) and Frail et al. (2001), the opening angle of the jet can be expressed as:

\begin{equation}
 \theta = 0.1~t_{\rm j,d}^{3/8}~(1+z)^{-3/8}~E_{\rm K, iso, 52}^{-1/8}~n_{0}^{1/8}
 \label{equ:theta}
\end{equation} 

\noindent where $t_{\rm j,d}$ is expressed in (observer-frame) days, and $E_{\rm K, iso, 52}$ is the isotropic kinetic energy of the fireball in units of $10^{52}$ erg, and $n_{0}$ is the circumburst density, where we have adopted the value $n_{0}=1$ cm$^{-3}$.  The LCs in Fig. \ref{fig:LC_decomp} are well described by a single power-law, with no evidence of a break up to the time when the SN starts to become the dominant source of light.  Thus a reasonable lower-limit to the break-time of $t_{\rm j,d}>5$.  To estimate the isotropic kinetic energy of the fireball, we used the equation for the radiative efficiency, but substituting isotropic values into it instead of beaming-corrected values.  Using a value of $E_{\gamma, \rm iso, 52} = 0.35$, for $\eta_{\gamma} =~ $0.1, 0.9, we find $E_{\rm K,52} =~$0.035--3.5.  Putting these values into equation \ref{equ:theta} gives opening angles in the range 8--14$^{\circ}$.  In comparison, Fong et al. (2012) found for a sample of LGRBs a range of opening angles of $\sim2^{\circ}-20^{\circ}$, with a median of $7^{\circ}$, and the lower-liming found here for GRB~140606B is commensurate with this range.  Therefore, using an opening angle of $\theta=8^{\circ}$, and values for the radiative efficiency of $\eta_{\gamma}=~$0.1--0.9, we derive the beaming-corrected kinetic energy of the fireball to be in the range $E_{\rm K} = (0.038$--3.1)$\times10^{50}$ erg.  The wide range (2 orders of magnitude) of the calculated kinetic energy arises primarily from the conservative range of radiative efficiencies considered here.

Next, we need an estimate of the velocity of the relativistic ejecta at $t-t_{0}=1$ days.  To do this we have used the following equation from Panaitescu \& Kumar (2000) for the propagation of the fireball in a homogeneous medium:

\begin{equation}
 \Gamma(t) = 6.3~E_{53}^{1/8}~n_{0}^{-1/8}~t_{\rm d}^{-3/8}
\end{equation}

\noindent where $E_{53}$ is the isotropic $\gamma$-ray energy in units of $10^{53}$ erg.  Using again $n_{0}=1$ cm$^{-3}$, we find $\Gamma=4.2$.  For $n=10$ and 0.1 cm$^{-3}$, we find $\Gamma=3.2,5.6$, respectively.

If GRB~140606B occurred in a wind-like medium, we also need to include an estimate of the expected pre-explosion mass-loss.  Typical values in the literature point to mass-loss rates of the order $\dot{M}=0.4-1.0\times10^{-5}$~M$_{\odot}$ yr$^{-1}$, with a wind velocity $\sim10^{3}$ km s$^{-1}$ (e.g. Margutti et al. 2013).  The equation for the propagation of the fireball in a wind-like medium is then (Panaitescu \& Kumar 2000):

\begin{equation}
 \Gamma(t) = 7.9~E_{53}^{1/4}~A_{*}^{-1/4}~t_{\rm d}^{-1/4}
\end{equation}

\noindent where $A_{*} = (\dot{\rm M}/10^{-5}$ M$_{\odot}$ yr$^{-1}$)/($v$/$10^{3}$ km s$^{-1}$).  For $A_{*} = 0.4,1.0$, we find $\Gamma=4.4,3.5$, respectively.  Therefore, regardless if the environment around the pre-explosion progenitor star is homogeneous or wind-like, the value of the bulk Lorentz factor at $t-t_{0}=1$ day is in the range 3--6.  The value of $\beta\approx0.95$ for this range, which implies that $\Gamma\beta\approx3-6$, which is the range of values used here.

The position of GRB~140606B in the $E_{\rm K}$--$\Gamma \beta$ plane is shown in Fig. \ref{fig:EK_GammaBeta_plane} as a shaded region, which accounts for the various assumptions used when we derived its high-energy properties.  The other data in the plot are from Margutti et al. (2014), and references therein.   Even allowing for the conservative assumptions, it is seen that the area occupied by GRB~140606B is between those occupied by SBO-GRBs and jetted-GRBs.  The power-law index to the top and bottom of the shaded box is $E_{\rm K} \propto (\Gamma\beta)^{-1.0}, (\Gamma\beta)^{-2.1}$, respectively.  Certainly the latter value is entirely consistent with those of the relativistic SNe IcBL and the SBO-GRBs, while the former is intermediate between the jetted-GRBs and SBO-GRBs.  However, an index of -1.0 is still much steeper than expected for a long-lived central engine, implying that the central engine powering the $\gamma$-ray emission was short-lived.  Therefore, using \textit{just} this line of reasoning, GRB~140606B is intermediate between SBO-GRBs and jetted-GRBs, and cannot be unambiguously classified as just one or the other type.

\subsection{The verdict}

Using the theoretical framework derived by Nakar \& Sari (2012), Singer et al. (2015) calculated a SBO radius of GRB~140606B of $\sim 10^{3}$ R$_{\odot}$, which is similar to the SBO radii calculated for $ll$GRBs 060218 (Campana et al. 2006) and 100316D (Starling et al. 2010).  This led Singer et al. (2015) to tentatively suggest that the $\gamma$-ray emission of GRB~140606B may also have arisen from the breakout from a dense wind and \textit{not} from the stellar surface.  In the scenario proposed by Margutti et al. (2015), and again recently by Nakar (2015), it is possible that instead of breaking out from a dense wind, that it did so from an extended low-mass envelope surrounding the pre-explosion progenitor star.  The origin of the extended low-mass envelope is poorly understood, but it is possible that it consists of material that was stripped from the surface of the pre-explosion progenitor star, but had not yet been ejected into space.   As the progenitors of GRB-SNe are thought to either be massive stars who ejected their outer layers into space prior to explosion via stellar winds, or a combination of stellar winds and mixing within the star (e.g. Yoon \& Langer 2005; Woosley \& Heger 2006), or stripped away by a binary companion (e.g. Fryer et al. 2014), the proposed scenario has merit.

Another line of evidence for a possible SBO origin is the shape and smoothness of the $\gamma$-ray LC for a given GRB-SN.  As reported by Bromberg et al. (2011), a key observable of $ll$GRBs are their smooth, non-variable $\gamma$-ray LCs compared to the more erratic $\gamma$-ray LCs of jetted-GRBs.  As noted by Burns (2014), the Fermi-GBM $\gamma$-ray LC of GRB~140606B consisted of a single long peak with a noisy tail, and displayed no signs of variability.\footnote{\raggedright \url{http://heasarc.gsfc.nasa.gov/FTP/fermi/data/gbm/triggers/2014/bn140606133/quicklook/glg_lc_all_bn140606133.gif}}  Thus the shape of the $\gamma$-ray LC of GRB~140606B lends some support to a SBO origin.

Along these lines, Nakar \& Sari (2012) suggested that a SBO is likely present in \textit{all} LGRB events, but due to its lower energy release relative to the prompt emission, it is not likely to be detected at redshifts exceeding $z\approx0.1$.  They do note that in this situation, the SBO may be in the form of a short pulse of photons with energies $>1$ MeV.  The rest-frame, peak/cutoff energy of $E_{p} \approx 800$ keV certainly approaches this value.  

If the $\gamma$-ray emission of GRB~140606B arose solely from a SBO, its isotropic energy $E_{\rm \gamma, iso, rest} \approx 4 \times 10^{51}$ erg is roughly 2--4 orders of magnitude greater than that observed for other suspected SBO-GRBs ($E_{\rm \gamma, iso, rest} \approx 10^{47}-5\times10^{49}$ erg; see Table \ref{table:bolo_compare}).  While the duration of its $\gamma$-ray emission is consistent with a SBO (Nakar \& Sari 2012), its isotropic-equivalent high-energy emission is suspiciously large, leading us to doubt whether GRB~140606B arose purely from a SBO.  

Conversely, if GRB~140606B arose purely from a jet, we are at odds as how to explain its presence as an outlier in the $E_{\rm p}$--$E_{\rm iso, \gamma}$ plane.  Over the years many authors have closely scrutinized the Amati relation, with opinions swinging back and forth as to whether it reflects a physical origin, or is simply due to selection effects.  For example, Nakar \& Piran (2005) concluded that roughly a quarter of all \emph{BATSE} bursts were inconsistent with the Amati relation.  They also noted that there was an absence of soft, bright bursts detected by \emph{BATSE}, \emph{BeppoSAX} and \emph{HETE-II}, and predicted that \emph{Swift} will detect hard, weak bursts.  GRB~140606B is clearly such a burst.

Band \& Preece (2005) also found similar results to Nakar \& Piran (2005), finding that 88\% of all \emph{BATSE} bursts were inconsistent with the Amati relation.  They concluded that the Amati and Ghirlanda (Ghirlanda et al. 2004) relations may be selection effects of the burst sample in which they were discovered, and that these selection effects may favour subpopulations for which these relations were valid.  However, Schaefer \& Collazzi (2007) argued that the Amati relation does in fact hold when only GRBs with spectroscopically determined redshifts were considered.

Collazzi et al. (2012) demonstrated that GRBs observed by \emph{BATSE}, \emph{Swift}, Suzaku and Konus all greatly violated the Amati relation, regardless if it had a spectroscopic redshift or not.  Each satellite had its own, greatly different, distribution in the $E_{\rm p}$--$E_{\rm iso, \gamma}$ plane, and these different distributions were dominated by selection effects.  Each satellite has a different trigger threshold, and a different threshold for the burst to obtain a measured $E_{\rm p, obs}$, which combined to make a diagonal cutoff.  For selection effects due to the intrinsic properties of the underlying burst population, the distribution of $E_{\rm p, obs}$ makes bursts with very low and very large values quite rare.  As such, for a detector with a high threshold, the combination of selection effects serves to allow only bursts within a region along the Amati relation (for that satellite) to be measured.   Similar conclusions were also reached by Shahmoradi \& Nemiroff (2010; 2011), who argued that the Amati and Ghirlanda relations can be considered to be real, but are heavily influenced by the detection thresholds of each GRB detector.

On the other hand however, it has been demonstrated by several authors that the time-integrated Amati and Ghirlanda relations are not due to selection effects (e.g. Ghirlanda 2011).  For example, Ghirlanda et al. (2010) analyzed the time-resolved spectra of a sample of 10 GRBs observed by Fermi, all of which had determined redshifts.  Their spectral evolution demonstrated that the peak energy tracks the emitted flux, and a strong time-resolved $E_{\rm p}$--$L_{\rm iso}$ correlation (e.g. Firmani et al. 2006) was present for individual GRBs -- a time-resolved correlation that was similar to the time-integrated correlation.  This result was independent of the spectral model used to fit the time-resolved spectra.  A similar analysis by Firmani et al. (2009) found similar conclusions.  These analyses all indicate that the Amati and Ghirlanda relations arise from physical origins and are not due to selection effects.

In summary:

\begin{itemize}
 \item GRB~140606B is an outlier in the Amati relation by 2-$\sigma$, and occupies the same region as nearby $ll$GRBs.  Taken at face value, this implies a possible SBO origin.
 \item The rest-frame peak energy measured for GRB~140606B is $E_{\rm p} \approx 800$ keV, which approaches the value expected for photons accelerated by a SBO ($>1$~MeV).  This also implies a possible SBO origin.
 \item GRB~140606B occupies the same region in the $M_{V, \rm p}$--$L_{\rm iso, \gamma}$ plane as jetted-GRBs, and is more than two orders of magnitude brighter in its $\gamma$-ray luminosity than $ll$GRBs.  This implies a jetted origin.
 \item In the $E_{\rm K}$--$\Gamma \beta$ plane, GRB~140606B occupies an intermediate region between $ll$GRBs (and relativistic SNe IcBL) and jetted-GRBs.  This is primarily due to the fact that radio and X-ray data are sorely missing, meaning that we were unable to tightly constrain the kinetic energy of the relativistic ejecta in this event (we were only able to constrain it to within two orders of magnitude).  Thus its placement in this plane does not allow us to unambiguously identify it as either a SBO-GRB or a jetted-GRB. 
\end{itemize}

Therefore, as was reported for the intermediate GRB~120422A (Schulze et al. 2014), GRB~140606B also has properties consistent with both SBO-GRBs \textit{and} jetted-GRBs.

\section{Discussion - Are there correlations between a GRB-SN's $\gamma$-ray emission and the SN's bolometric properties?}
\label{section:discussion}


\begin{figure*}
 \centering
 \includegraphics[bb=0 0 695 491, scale=0.65]{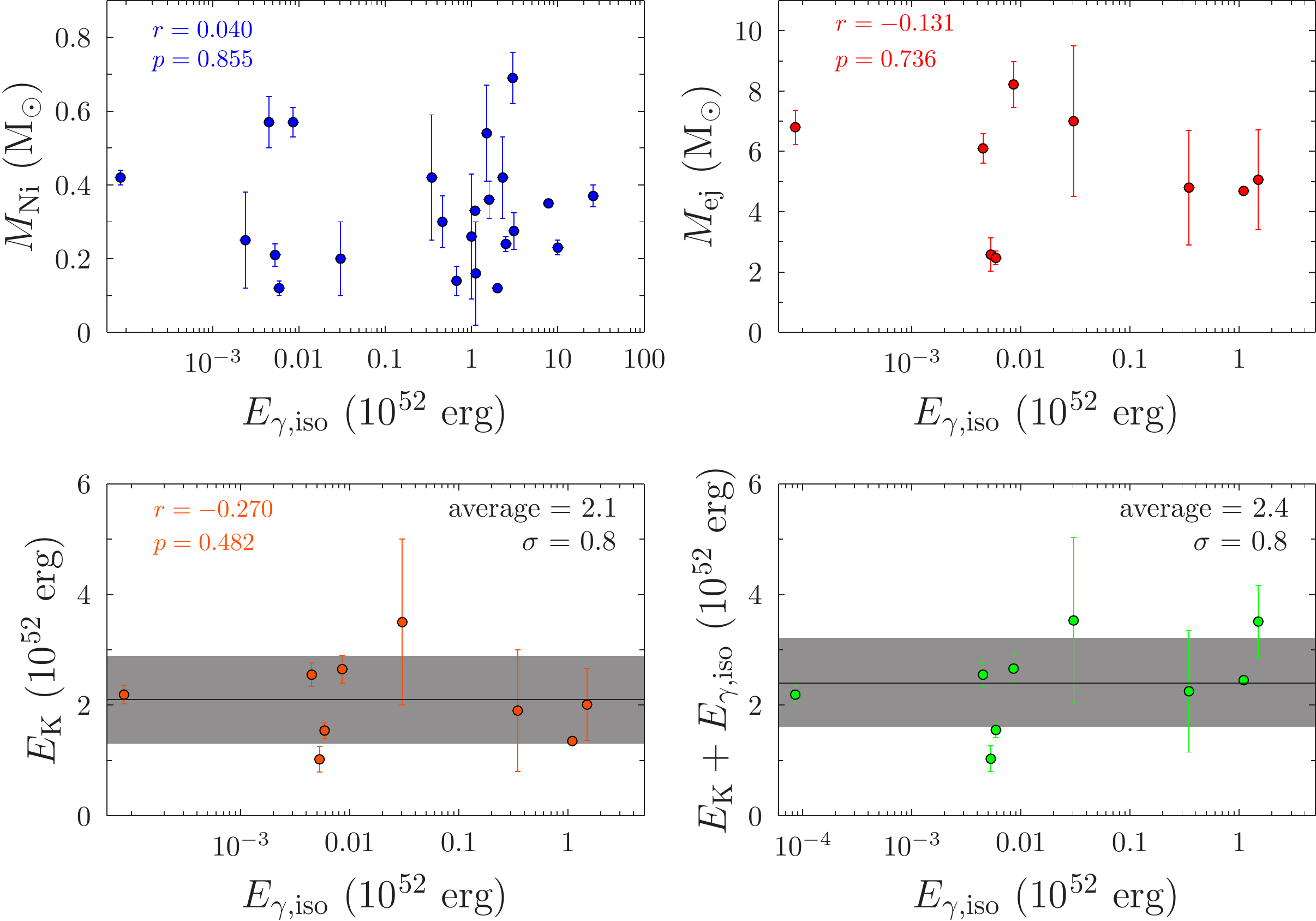}
 \caption{A comparison of the $\gamma$-ray energetics $vs$ bolometric properties of a sample of GRB-SNe (see Table \ref{table:bolo_compare}).  No statistically significant correlation is seen between $E_{\rm \gamma, iso}$ and $M_{\rm Ni}$, $M_{\rm ej}$ or $E_{\rm K}$ (the latter two values have been determined only for those GRB-SNe that have an estimate of its peak photospheric velocity as determined from spectroscopy of the SN itself, e.g. C13, Lyman et al. 2014).  The average kinetic energy is $2.1\times10^{52}$ erg, with a standard deviation of $\sigma=0.8\times10^{52}$ erg, which are shown as the shaded area in the bottom left panel.  It is also seen that the sum of $E_{\rm \gamma, iso} + E_{\rm K}$ for both GRB- and $ll$SNe has an average value of $2.4\times10^{52}$ erg, with a standard deviation of $\sigma=0.8\times10^{52}$ erg, which is shown as the shaded grey region in the bottom right panel.  The average kinetic energy and average summed energies are consistent with each other, and they provide constraints on the explosion mechanism of GRB-SNe.}
 \label{fig:bolo_compare}
\end{figure*}

We gathered published data for a sample of GRB-SNe in order to compare the isotropic high-energy $\gamma$-ray emission with the bolometric properties of the associated SN.  The data are presented in Table \ref{table:bolo_compare}, and the comparisons displayed in Fig. \ref{fig:bolo_compare}.  The Pearson's correlation coefficient ($r$) and the two-point probability of a chance correlation ($p$) are also displayed for each comparison.  We have only plotted ejecta masses and kinetic energies for those events where an estimate of the peak photospheric velocity was directly determined from spectroscopy of the SN itself.  It is seen that no statistically significant correlation is seen between any of the measured properties, which is reflected in the value of $r$ in each comparison, which are roughly consistent with $r=0$.

It is seen that the average kinetic energy of the GRB-SN sample is $\bar{E}_{\rm K}=2.1\times 10^{52}$ erg, with a standard deviation of $\sigma=0.8\times 10^{52}$ erg.  This is entirely consistent with the median value of the kinetic energy found for $N=20$ GRB-SNe in C13 ($\tilde{E}_{\rm K} = 2.2 \times 10^{52}$ erg; $\sigma=1.5 \times 10^{52}$~erg).  We also computed the total energetics of a GRB-SN event, i.e. $E_{\rm \gamma, iso} + E_{\rm K}$, finding an average value of $2.4\times 10^{52}$ erg, with a standard deviation of $\sigma=0.8\times 10^{52}$ erg.  Every one of the considered GRB-SNe are within this range, with the exception of SN~2006aj, which falls below the lower-limit.  Indeed the value of the kinetic energy used here for SN~2006aj is larger than that used in other works, (e.g. Pian et al. 2006 found $E_{\rm K} \approx 0.6 \times 10^{52}$ erg), meaning it would be even more of an outlier.

The nature of the central engine of GRB-SNe is still debated, where the consensus is split between magnetars and accreting BHs.  Recent work, e.g. Mazzali et al. (2014), found that, when the modelling of GRB-SNe properly considers the asphericity of the ejecta, the total kinetic energy of all GRB-SNe clusters around $10^{52}$ erg, with an upper limit of $2\times10^{52}$ erg.  The caveat of the data presented in Table \ref{table:bolo_compare} is that they have been modelled with an analytical model that assumes that the ejecta is spherically symmetric.  As noted by Mazzali et al. (2014), when asphericity is included in hydrodynamical models coupled with radiative transfer simulations, the corresponding kinetic energies are smaller by a factor of $\approx 2-5$ (see as well Cano et al. 2014).  

The average total energy of the relativistic and non-relativistic components have implications for the total energy budget available in a GRB-SN event.  Proponents of the magnetar central engine point out that the total energy available in the central engine to power the SN cannot be in excess of $2\times10^{52}$ erg (e.g. Thompson et al. 2004; Metzger et al. 2011).  The total amount of energy available in the collapsar model is less certain (see Woosley \& Bloom 2006), and depends on uncertain mechanisms for turning disk binding energy or black hole rotation energy into directed relativistic outflows (which can be neutrinos, magnetic instabilities in the accretion disk, and MHD extraction of rotational energy from the BH).  Larger kinetic energies can be obtained via MHD origins, and can in theory be in excess of $2\times10^{52}$ erg, however a precise upper limit is harder to precisely determine.  In conclusion, the values presented here are fully consistent with the result of Mazzali et al. (2014), who suggest that the central engine of all GRB-SNe may indeed be a magnetar, but an accreting BH central engine cannot be ruled out.

\section{Conclusions}
\label{section:conclusions}

In this paper we presented optical and NIR photometry and spectroscopy of GRB~140606B and its accompanying SN.  The brightness of the SN is commensurate with those of other GRB-SNe in terms of peak, rest-frame $V$-band brightness (only $\approx 0.1$ mag brighter than SN~1998bw).  The bolometric properties of the SN ($M_{\rm Ni} = 0.4\pm0.2$ $\rm M_{\odot}$, $M_{\rm ej} = 5\pm2$ $\rm M_{\odot}$, $E_{\rm K} = 2\pm1 \times 10^{52}$ erg), which were determined using the method presented in C13 (and used also in Xu et al. 2013; Cano et al. 2014) are all similar to those of the general GRB-SN population.

Spectroscopically, the SN is typical of other GRB-SNe: it displays broad, blended features in the four-epoch time-series, and has a peak photospheric velocity (determined using blueshifted Fe \textsc{ii} $\lambda$5169 as a proxy) of $v_{\rm ph} \approx 20,000$ km s$^{-1}$.  This velocity is very typical of that measured for other GRB-SNe (C13 found that the average peak photospheric velocity of a sample of GRB-SNe to be  $v_{\rm ph} \approx 20,000$ km s$^{-1}$) and non-GRB-SNe (Lyman et al.:  $v_{\rm ph} \approx 19,100$ km s$^{-1}$).  Thus, both in terms of the SN's bolometric properties, and the value of the peak photospheric velocity, make the SN associated with GRB~140606B quite a typical GRB-SN. 

Next, it was found that the SN accompanying GRB~140606B also follows the $k$--$s$ relation found for a sample of GRB-SNe (which also included the engine-drive SN~2009bb; Pignata et al. 2011, Soderberg et al. 2010) by C14.  This implies that this SN can also be used to constrain the fundamental parameters of cosmological models, as done in CJ14 and Li et al. 2014.  

We attempted to constrain the rest-frame extinction by modelling contemporaneous X-ray and optical data of GRB~140606B.  However, the poorly sampled X-ray LC meant we were not able to constrain the rest-frame extinction very precisely, finding a value $E(B-V)_{\rm rest} = 0.16\pm0.14$ mag.

The SFR of the host galaxy of GRB~140606B is moderately small,  ($\approx 0.05$ M$_{\odot}$ yr$^{-1}$), but not abnormally so, and indeed similar to values measured for the hosts of other GRB-SNe (e.g. Savaglio et al. 2009; Kr\"uhler et al. 2015).  It was seen that the host of GRB~140606B has a companion that is a projected distance of only $\sim 5$ kpc.  The SFR of the companion galaxy is similar to that of the GRB host, $\approx$~0.07~M$_{\odot}$~yr$^{-1}$.

We also investigated the idea presented in Singer et al. (2015) that GRB~140606B may be a SBO GRB-SN rather than a jet-driven GRB-SN.  Motivation for this arises from the fact that GRB~140606B is an outlier in the Amati relation, and occupies the same region in the $E_{\rm p}$--$E_{\rm iso, \gamma}$ plane as low-luminosity GRBs 980425, 031203, 100316D and intermediate-GRB 120422A.  Additionally, the value of the peak, rest-frame energy was found to be $\approx 800$ keV, which is close to, albeit less than that expected for photons accelerated by a SBO, which are expected to have energies in excess of 1 MeV.  These two observations indicate a possible SBO origin for GRB~140606B.  However, the placement of the SN associated with GRB~140606B occupies the same place in the $M_{V, \rm p}$--$L_{\rm iso, \gamma}$ plane as jetted-GRBs, and is two orders of magnitude brighter than that measured for $ll$GRBs.  This observation indicates a possible jetted-origin.  Finally, its position in the $E_{\rm K}$--$\Gamma \beta$ was loosely constrained due to the lack of radio of X-ray observations, which limited how precisely we were able to determine the kinetic energy of the relativistic ejecta in this event (which spanned two orders of magnitude).  It was seen that its placement in this plane was intermediate between SBO-GRBs and jetted-GRBs, and not unambiguously associated with either group.  We were forced to conclude that, in a similar fashion to the intermediate GRB~120422A, GRB~140606B has high-energy and SN properties that are consistent with both SBO-GRBs and jetted-GRBs.

Finally, we search for correlations between the isotropic $\gamma$-ray emission and the bolometric properties of GRB-SNe, finding that no statistically significant correlation is present.   It was seen that the sum of $E_{\rm \gamma, iso} + E_{\rm K}$ for our sample of GRB-SNe has an average value of $2.4\times10^{52}$~erg, with a standard deviation of $\sigma=0.8\times10^{52}$~erg.  All of the GRB-SNe in our sample, with the exception of SN~2006aj, are within this range (Fig. \ref{fig:bolo_compare}), which has implications for the total energy budget available to power both the relativistic and non-relativistic components of a GRB-SN event.  This energy limit is consistent with a magnetar central engine for GRB-SNe, but it does not rule out accreting BHs, where the total amount of energy available to power a SN is less certain.

\begin{table*}
\scriptsize
\centering
\setlength{\tabcolsep}{5pt}
\setlength{\extrarowheight}{3pt}
\caption{Physical Properties of GRB-SNe}
\label{table:bolo_compare}
\begin{tabular}{ccccccccc}
\hline							
GRB	&	Type	&	$E_{\rm \gamma, iso}$ ($10^{52}$ erg)	&		$M_{\rm Ni}$ (M$_{\odot}$)				&		$M_{\rm ej}$ (M$_{\odot}$)				&		$E_{\rm K}$ ($10^{52}$ erg)				&	$E_{\rm \gamma, iso}+E_{\rm K}$ ($10^{52}$ erg)	&	$E_{\rm K}/E_{\rm \gamma, iso}$	&	Ref.	\\
\hline					
980425	&	XRF	&	0.000086	&	$	0.42	\pm	0.02	$	&	$	6.8	\pm	0.6	$	&	$	2.2	\pm	0.2	$	&	2.2	&	25550.0	&	(1,2)	\\
030329	&	GRB	&	1.5	&	$	0.54	\pm	0.13	$	&	$	5.1	\pm	1.7	$	&	$	2.0	\pm	0.7	$	&	3.5	&	1.3	&	(1,2)	\\
031203	&	XRF	&	0.0086	&	$	0.57	\pm	0.04	$	&	$	8.2	\pm	0.8	$	&	$	2.7	\pm	0.3	$	&	2.7	&	309.2	&	(1,2)	\\
060218	&	XRF	&	0.0053	&	$	0.21	\pm	0.03	$	&	$	2.6	\pm	0.6	$	&	$	1.0	\pm	0.2	$	&	1.0	&	192.5	&	(1,2)	\\
091127	&	GRB	&	1.1	&	$	0.33	\pm	0.01	$	&	$	4.7	\pm	0.1	$	&	$	1.4	\pm	0.0	$	&	2.5	&	1.2	&	(1,2)	\\
100316D	&	XRF	&	0.0059	&	$	0.12	\pm	0.02	$	&	$	2.5	\pm	0.2	$	&	$	1.5	\pm	0.1	$	&	1.5	&	261.0	&	(1,2)	\\
120422A	&	Inter	&	0.0045	&	$	0.57	\pm	0.07	$	&	$	6.1	\pm	0.5	$	&	$	2.6	\pm	0.2	$	&	2.6	&	566.7	&	(1,3)	\\
130702A	&	Inter	&	0.0304	&	$	\approx 0.2			$	&	$	\approx 7			$	&	$	\approx 3.5			$	&	3.5	&	115.1	&	(4)	\\
140606B	&	GRB	&	0.347	&	$	0.42	\pm	0.17	$	&	$	4.8	\pm	1.9	$	&	$	1.9	\pm	1.1	$	&	2.2	&	5.5	&	this work	\\
\ldots	&	\ldots	&	\ldots	&	$	\ldots			$	&	$	\ldots			$	&	$	\ldots			$	&	\ldots	&	\ldots	&	\ldots	\\
990712	&	GRB	&	0.67	&	$	0.14	\pm	0.04	$	&	$	\ldots			$	&	$	\ldots			$	&	\ldots	&	\ldots	&	(1,2)	\\
011121	&	GRB	&	7.8	&	$	0.35	\pm	0.01	$	&	$	\ldots			$	&	$	\ldots			$	&	\ldots	&	\ldots	&	(1,2)	\\
020405	&	GRB	&	10.0	&	$	0.23	\pm	0.02	$	&	$	\ldots			$	&	$	\ldots			$	&	\ldots	&	\ldots	&	(1,2)	\\
020903	&	XRF	&	0.0024	&	$	0.25	\pm	0.13	$	&	$	\ldots			$	&	$	\ldots			$	&	\ldots	&	\ldots	&	(1,2)	\\
021211	&	GRB	&	1.12	&	$	0.16	\pm	0.14	$	&	$	\ldots			$	&	$	\ldots			$	&	\ldots	&	\ldots	&	(1,2)	\\
041006	&	GRB	&	3.0	&	$	0.69	\pm	0.07	$	&	$	\ldots			$	&	$	\ldots			$	&	\ldots	&	\ldots	&	(1,2)	\\
050525A	&	GRB	&	2.5	&	$	0.24	\pm	0.02	$	&	$	\ldots			$	&	$	\ldots			$	&	\ldots	&	\ldots	&	(1,2)	\\
050824	&	GRB	&	1.0	&	$	0.26	\pm	0.17	$	&	$	\ldots			$	&	$	\ldots			$	&	\ldots	&	\ldots	&	(1,2)	\\
060729	&	GRB	&	1.6	&	$	0.36	\pm	0.05	$	&	$	\ldots			$	&	$	\ldots			$	&	\ldots	&	\ldots	&	(1,2)	\\
060904B	&	GRB	&	2.0	&	$	0.12	\pm	0.01	$	&	$	\ldots			$	&	$	\ldots			$	&	\ldots	&	\ldots	&	(1,2)	\\
090618	&	GRB	&	25.7	&	$	0.37	\pm	0.03	$	&	$	\ldots			$	&	$	\ldots			$	&	\ldots	&	\ldots	&	(1,2)	\\
120729A	&	GRB	&	2.3	&	$	0.42	\pm	0.11	$	&	$	\ldots			$	&	$	\ldots			$	&	\ldots	&	\ldots	&	(2,5)	\\
130215A	&	GRB	&	3.1	&	$	0.28	\pm	0.05	$	&	$	\ldots			$	&	$	\ldots			$	&	\ldots	&	\ldots	&	(2,5)	\\
130831A	&	GRB	&	0.46	&	$	0.30	\pm	0.07	$	&	$	\ldots			$	&	$	\ldots			$	&	\ldots	&	\ldots	&	(2,5)	\\
\hline							
\end{tabular}
\begin{flushleft}
NB: Ejecta masses ($M_{\rm ej}$) and kinetic energies ($E_{\rm K}$) are calculated only for GRB-SNe where a peak photospheric velocity has been determined from spectroscopy of the SN.\\ 
$\dagger$: D'Elia et al. (2015) did not estimate the errors on their bolometric properties.  We have assumed errors of $M_{\rm Ni} = 0.20\pm0.10$~M$_{\odot}$, $M_{\rm ej} = 7.0\pm3.0$~M$_{\odot}$ and $E_{\rm K} = (3.5\pm1.5)\times10^{52}$ erg.\\
Refs: (1) C13; (2) Hjorth \& Bloom (2012); (3) Schulze et al. (2014); (4) D'Elia et al. (2015); (5) Cano et al. (2014) \\
\end{flushleft}
\end{table*}


\section*{Acknowledgments}

We thank Giorgos Leloudas for his expertise in analysing the optical spectra of the SN.  We also thank Y. Cao for carrying out the Keck observations.

ZC is funded by a Project Grant from the Icelandic Research Fund.  PJ and MF acknowledge support by the University of Iceland Research fund.  JPUF acknowledges support from the ERC-StG grant EGGS-278202.  The Dark Cosmology Centre is funded by the DNRF.  S. Schulze acknowledges support from CONICYT-Chile FONDECYT 3140534, Basal-CATA PFB-06/2007, and Project IC120009 ``Millennium Institute of Astrophysics (MAS)´´ of Iniciativa Cient\'{\i}fica Milenio del Ministerio de Econom\'{\i}a, Fomento y Turismo.



\label{lastpage}

\end{document}